\documentclass{osa-article}

\journal{osac}

\usepackage{rotating}
\usepackage{titling}
\usepackage{pdflscape}
\usepackage{adjustbox}
\usepackage{textcomp}
\articletype{Research Article}

\usepackage{scalerel}
\usepackage{tikz}
\usetikzlibrary{svg.path}

\definecolor{lime}{HTML}{A6CE39}

\DeclareRobustCommand{\orcidicon}{%
    {\mbox{\scalerel*{
    \tikzset{
      orcidlogo/.pic={
        \fill[lime] svg{M256,128c0,70.7-57.3,128-128,128C57.3,256,0,198.7,0,128C0,57.3,57.3,0,128,0C198.7,0,256,57.3,256,128z};
        \fill[white] svg{M86.3,186.2H70.9V79.1h15.4v48.4V186.2z}
                     svg{M108.9,79.1h41.6c39.6,0,57,28.3,57,53.6c0,27.5-21.5,53.6-56.8,53.6h-41.8V79.1z M124.3,172.4h24.5c34.9,0,42.9-26.5,42.9-39.7c0-21.5-13.7-39.7-43.7-39.7h-23.7V172.4z}
                     svg{M88.7,56.8c0,5.5-4.5,10.1-10.1,10.1c-5.6,0-10.1-4.6-10.1-10.1c0-5.6,4.5-10.1,10.1-10.1C84.2,46.7,88.7,51.3,88.7,56.8z};
      }
    }
    \begin{tikzpicture}[yscale=-1,transform shape]
    \pic{orcidlogo};
	\end{tikzpicture}
	\hspace{-2mm}
	}{|}}}
}

\foreach \x in {A, ..., Z}{%
	\expandafter\xdef\csname orcid\x\endcsname{\noexpand\href{https://orcid.org/\csname orcidauthor\x\endcsname}{\noexpand\orcidicon}}
}

\usepackage{hyperref}

\begin{document}

\title{The vector-apodizing phase plate coronagraph: design, current performance, and future development}
\author{D. S. Doelman,\authormark{1}* \orcidA{}
F. Snik,\authormark{1} \orcidP{}
E. H. Por,\authormark{1,2} \orcidB{}
S. P. Bos,\authormark{1}
G.P.P.L. Otten,\authormark{1,3} \orcidC{}
M. Kenworthy,\authormark{1} \orcidL{}
S. Y. Haffert,\authormark{1,4} \orcidD{}
M. Wilby,\authormark{1}
A.~J.~Bohn,\authormark{1} \orcidE{}
B. J. Sutlieff,\authormark{5,1} \orcidF{}
K. Miller,\authormark{1} \orcidM{}
M. Ouellet,\authormark{1,6} 
J. de Boer, \authormark{1}
C. U. Keller,\authormark{1} \orcidN{}
M. J. Escuti,\authormark{7} \orcidK{}
S. Shi,\authormark{7}
N.Z. Warriner,\authormark{7}
K. J. Hornburg, \authormark{7}
J. L. Birkby,\authormark{8,5,9} \orcidO{}
J. Males,\authormark{4}
K. M. Morzinski,\authormark{4}
L.M. Close,\authormark{4}
J. Codona,\authormark{4}
J. Long,\authormark{4}
L. Schatz,\authormark{4}
J. Lumbres,\authormark{4,10}
A. Rodack,\authormark{4}
K. Van Gorkom,\authormark{4}
A Hedglen,\authormark{4}
O. Guyon,\authormark{4,10,11,12}
J. Lozi,\authormark{11}
T. Groff,\authormark{13}
J. Chilcote,\authormark{14}
N. Jovanovic,\authormark{15} \orcidG{}
S. Thibault,\authormark{6}
C. de Jonge,\authormark{16}
G. Allain,\authormark{6}
C. Vall\'ee,\authormark{6}
D. Patel,\authormark{6}
O. C\^ot\'e,\authormark{6}
C. Marois,\authormark{17,18}
P. Hinz,\authormark{4,19}
J. Stone,\authormark{4,20}
A. Skemer,\authormark{19} \orcidQ{}
Z. Briesemeister,\authormark{19}
A. Boehle,\authormark{21} \orcidH{}
A. M. Glauser,\authormark{21}
W. Taylor,\authormark{22} 
P. Baudoz,\authormark{23}
E. Huby, \authormark{23}
O. Absil,\authormark{24} \orcidJ{}
B. Carlomagno,\authormark{24}
and C. Delacroix\authormark{24} \orcidI{}}

\address{\authormark{1} Leiden Observatory, Leiden University, P.O. Box 9513, 2300 RA Leiden, The Netherlands\\
\authormark{2} Space Telescope Science Institute, 3700 San Martin Drive, Baltimore, MD 21218, USA\\
\authormark{3} Aix Marseille Univ, CNRS, CNES, LAM, Marseille, France \\
\authormark{4} Steward Observatory, 933 North Cherry Avenue, University of Arizona, Tucson, AZ 85721, USA\\
\authormark{5} Anton Pannekoek Institute for Astronomy, University of Amsterdam, Science Park 904, 1098 XH Amsterdam, The Netherlands\\
\authormark{6} COPL, D\'epartement de physique, g\'enie physique et optique, Universit\'e Laval, 2375, rue de la Terrasse, Quebec,G1V 0A6, Canada.\\
\authormark{7} Department of Electrical and Computer Engineering, North Carolina State University, Raleigh, NC 27695, USA\\
\authormark{8} Astrophysics, Department of Physics, University of Oxford, Denys Wilkinson Building, Keble Road, Oxford, OX1 3RH, United Kingdom\\
\authormark{9} Harvard \& Smithsonian | Center for Astrophysics, 60 Garden St, Cambridge, MA 02138, USA\\ 
\authormark{10} College of Optical Sciences, University of Arizona, 1630 E University Blvd, Tucson, AZ 85719, USA\\
\authormark{11} Subaru Telescope, National Astronomical Observatory of Japan, National Institutes of Natural Sciences (NINS), 650 North A$\!$`oh\={o}k\={u} Place, Hilo, HI, 96720, U.S.A.\\
\authormark{12} Astrobiology Center of NINS, 2-21-1, Osawa, Mitaka, Tokyo, 181-8588, Japan \\
\authormark{13} NASA-Goddard Space Flight Center, Greenbelt, MD, USA \\
\authormark{14} Department of Physics, University of Notre Dame, South Bend, IN, USA \\
\authormark{15} Department of Astronomy, California Institute of Technology, 1200 East California Blvd., Pasadena, CA, 91125, U.S.A.\\
\authormark{16} SRON Netherlands Institute for Space Research, Landleven 12, 9747AD Groningen, The Netherlands \\ 
\authormark{17} National Research Council of Canada Herzberg, 5071 West Saanich Road, Victoria, BC V9E 2E7, Canada \\
\authormark{18} Department of Physics and Astronomy, University of Victoria, 3800 Finnerty Road, Victoria, BC V8P 5C2, Canada \\
\authormark{19} Department of Astronomy and Astrophysics, University of California, Santa Cruz, 1156 High Street, Santa Cruz, CA, 95064, U.S.A\\
\authormark{20} Naval Research Laboratory, Remote Sensing Division, 4555 Overlook Ave. SW, Washington, DC 20375, USA\\
\authormark{21} Institute for Particle Physics and Astrophysics, ETH Zurich, Wolfgang-Paulistr. 27, 8093 Zurich, Switzerland\\
\authormark{22} UK Astronomy Technology Centre, STFC, Blackford Hill, Edinburgh EH9 3HJ, UK \\
\authormark{23} LESIA, Observatoire de Paris, Universit\'e PSL, CNRS, Sorbonne Universit\'e, Universit\'e de Paris, 5 place Jules Janssen, 92195 Meudon, France\\
\authormark{24} STAR Institute, Universit\'e de Li\`ege, All\'ee du Six Ao\^ut 19c, B-4000 Li\`ege, Belgium  }

\email{\authormark{*}doelman@strw.leidenuniv.nl} 



\begin{abstract}
Over the last decade, the vector-apodizing phase plate (vAPP) coronagraph has been developed from concept to on-sky application in many high-contrast imaging systems on 8-m class telescopes. 
The vAPP is an geometric-phase patterned coronagraph that is inherently broadband, and its manufacturing is enabled only by direct-write technology for liquid-crystal patterns. 
The vAPP generates two coronagraphic PSFs that cancel starlight on opposite sides of the point spread function (PSF) and have opposite circular polarization states. 
The efficiency, that is the amount of light in these PSFs, depends on the retardance offset from half-wave of the liquid-crystal retarder.
Using different liquid-crystal recipes to tune the retardance, different vAPPs operate with high efficiencies ($>96\%$) in the visible and thermal infrared (0.55 \textmu m to 5 \textmu m). 
Since 2015, seven vAPPs have been installed in a total of six different instruments, including Magellan/MagAO, Magellan/MagAO-X, Subaru/SCExAO, and LBT/LMIRcam.
Using two integral field spectrographs installed on the latter two instruments, these vAPPs can provide low-resolution spectra (R$\sim$30) between 1 $\mu$m and 5 $\mu$m. 
We review the design process, development, commissioning, on-sky performance, and first scientific results of all commissioned vAPPs.
We report on the lessons learned and conclude with perspectives for future developments and applications.
\end{abstract}

\section{Introduction}
\begin{figure}
    \centering
    \includegraphics[width=\linewidth]{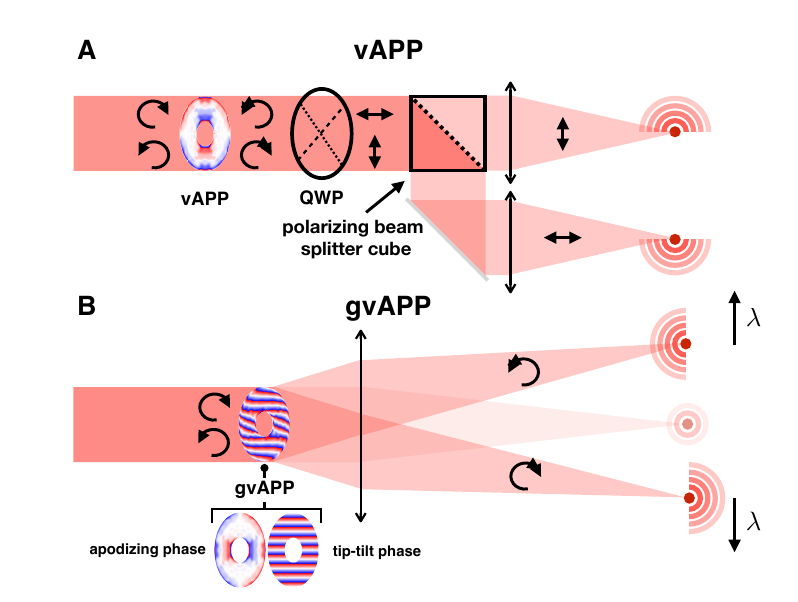}
    \caption{Schematic of two different vAPP implementations. \textit{Top:} The vAPP, where PSFs with dark zones on opposite sides and opposite polarization states are separated using a quarter-wave plate and a polarizing beam-splitter.  \textit{Bottom:} The gvAPP where polarization states are separated by adding a phase ramp (= polarization grating) to the vAPP phase pattern, which generates opposite tilt for light with left- and right-circular polarization, adapted from \cite{bos2018fully}.}
    \label{fig:APP_schematic}
\end{figure}

Many different coronagraphs have been proposed since the creation of the first solar coronagraph in 1939 by Bernard Lyot \cite{lyot1939study}. 
Originally, the existing coronagraph concepts could be organized in a family tree \cite{mawet2012review}. However, this tree was cut down in the Lorentz Center workshop paper \cite{ruane2018}, because new developments in coronagraph design lead to the merging of various branches.
Due to its simplicity, an adapted version of the Lyot coronagraph is still the most used coronagraph for ground-based high-contrast imaging systems.
The use of Lyot coronagraphs resulted in many scientific breakthroughs in understanding exoplanets and circumstellar disks.
Preliminary results of the two largest exoplanet surveys are presented by Nielsen et al. (2019) \cite{nielsen2019gemini} (GPIES) and Vigan et al. (2020) \cite{vigan2020sphere} (SHINE), and a summary paper on circumstellar disks was presented by Avenhaus et al. (2018) \cite{avenhaus2018disks}.\\
The Lyot coronagraph and other focal-plane coronagraphs require accurate centering of the star on their focal-plane masks.
Pupil-plane coronagraphs do not have this disadvantage, as the coronagraphic PSF does not change significantly with position in the field of view.
The Apodizing Phase Plate (APP) coronagraph is a single-optic pupil-plane coronagraph \cite{codona2006,kenworthy2007}.
An APP modifies the phase in the pupil plane to create regions in the PSF where the star light is suppressed, so-called dark zones. 
A pupil-plane coronagraph has some distinct advantages over a focal-plane coronagraph.
First, the coronagraph is simple because it consists of only a single optic, making it easy to install in any high-contrast imaging instrument. 
Second, the coronagraphic performance is insensitive to tip-tilt errors caused by vibrations or residual wavefront from the adaptive optics \textcolor{black}{(AO)} system.
Moreover, tip-tilt insensitivity is good for the near-infrared (3-5 $\mu$m) where nodding is required to remove the background.
With nodding the PSF moves over a large distance ($\gg\lambda/D$) on the detector and alignment with a focal plane mask afterwards takes time and the final position is not always the same, resulting in PSF differences. 
For a pupil-plane coronagraph no realignment is required after nodding, increasing the on-axis time. 
Third, the coronagraph design can easily be adapted to include complex pupil shapes (e.g. segments/spiders). 
Fourth, all objects have the coronagraphic PSF, enabling high-contrast imaging of binary systems. 
These advantages contributed to success of the APP, having imaged multiple substellar companions (e.g. $\beta$ Pictoris b, discovering HD100546~b and HD 984, \cite{quanz2010,Quanz13,meshkat2015}).
Disadvantages of pupil-plane coronagraphs are the high intensity of the stellar PSF often resulting in saturation, the lower planet throughput due to the reshaping of the PSF, and the larger inner working angle with respect to the current best focal-plane coronagraphs.
More specifically for the APP, the properties of the diamond-turned phase plate limited the exoplanet yield. 
As the APP applies dynamic (chromatic) phase, the APP performance is only optimal for a single wavelength, and the dark zone contrast is degraded for broadband light.
More importantly, the diamond turning requires smooth phase transitions, which limited the APP designs to an outer working angle of 9 $\lambda/D$ and restricted dark zones to a single side of the PSF.\\
The vector-Apodizing Phase Plate (vAPP) coronagraph is an upgraded version of the APP coronagraph. The vAPP induces geometric (or Pancharatnam-Berry) phase \cite{pancharatnam1956,berry1987} for circularly polarized light \cite{snik2012,otten2014}. 
The vAPP generates two coronagraphic PSFs that have dark zones on opposite sides of the point spread function and have opposite circular polarization states, see Fig. \ref{fig:APP_schematic}. 
The vAPP is a patterned half-wave retarder where the fast-axis orientation changes as a function of position. 
The induced phase, $\phi$, is equal to plus/minus twice the fast-axis orientation $\theta$: $\phi = \pm 2\theta$, with opposite sign for the opposite circular polarization states.
Geometric phase is by definition achromatic, and the efficiency, the percentage of light that acquires this phase, depends on the retardance offset from half-wave. 
\textcolor{black}{The fraction of the light that does not acquire the phase is called the ``polarization leakage''}. 
The broadband performance of the vAPP is therefore determined by the retardance as a function of wavelength.  \\
The vAPP is manufactured with liquid-crystal technology. 
A direct-write system is used to print the desired fast-axis orientation pattern in a liquid-crystal photo-alignment layer that has been deposited on a substrate \cite{miskiewicz2014}. 
The induced orientation depends on the linear polarization of the incoming light of the direct-write system. 
Multiple layers of self-aligning birefringent liquid-crystals are deposited on top with varying thickness and twist, carefully designed to generate the required half-wave retardance \cite{komanduri2012,komanduri2013}. 
The stack is referred to as a multi-layered twisted retarder (MTR). 
By tuning the twist and thickness of layers in the MTR, very high efficiencies  ($>96$\%) can be achieved for large wavelength ranges, e.g. $2-5 \mu$m \cite{otten2014}. \\
\textcolor{black}{Both unpolarized light and linearly-polarized light contain equal amounts of left- and right-circularly polarized light.} 
As opposite circular polarization states create a PSF with a dark hole on opposite sides, the vAPP implements circular polarization splitting to separate two complementary dark holes.
The relative intensity of the two coronagraphic PSFs depends on the circular polarization state of the incoming light.

A vAPP for broadband imaging is obtained in combination with a polarizing beamsplitter (e.g. a Wollaston prism) and a quarter-wave plate, but the non-coronagraphic PSFs corresponding to various polarization leakage terms degrade the contrast in the dark holes \cite{snik2012,otten2014,bos2018fully}.
The polarization leakage can be separated from the coronagraphic PSFs by adding a grating pattern (= phase tilt) to the phase pattern, i.e.~the ``grating-vAPP'' (gvAPP) \cite{otten2014a}. 
The gvAPP has the same advantages as all pupil-plane coronagraphs, and also overcomes \textcolor{black}{three} limitations of the vAPP. 
First, no polarization splitting or filtering optics are required due to the grating.
\textcolor{black}{The result is that the gvAPP is also unaffected by polarization crosstalk which could arise in the instrument between the vAPP and the polarization splitting.
Second, the polarization leakage of the gvAPP is physically separated from the coronagraphic PSFs. The performance of the gvAPP is therefore largely unaffected by the leakage term.
For the vAPP, the leakage term of the vAPP for unpolarized light would present a low-intensity ($\sim1\%$) non-coronagraphic PSF at the location of the coronagraphic PSF.
Moreover, for linearly polarized light this PSF would even be coherent with the coronagraphic PSF.
Both reduce the contrast at the smallest separations.}
Third, the gvAPP has two dark holes on opposite side, increasing the search space by a factor two.
This comes at the cost of a factor of two in exoplanet throughput, with half of the light being imaged at the bright side of one of the two coronagraphic PSFs. 
We note that for unknown companions this factor is the same when using an APP, having to rotate the APP to image both sides.
Advantages of the two vAPP dark holes are that they have the same AO performance, such that one PSF can be used as a reference for post processing \cite{otten2017}, and that the antisymmetric PSF is beneficiary for wavefront sensing \cite{bos2019}.
Moreover, the direct-write system is capable of writing much finer structures ($\sim 1 \mu$m) compared to diamond turning ($\sim 50 \mu$m) and has the ability to write discontinuous phase steps. 
These properties enable a tremendously increased phase pattern complexity that can be used for holography \cite{haffert2018} and manufacture more optimal phase patterns \cite{doelman2020}.\\
A third kind of vAPP is the double-grating vAPP (dgvAPP). 
The dgvAPP combines a gvAPP with a second polarization grating on a separate substrate \cite{doelman2020}. 
This second grating is identical to the gvAPP grating and has the opposite effect, diffracting both the main beams back on axis.
Polarization leakage of the liquid-crystal film on the first substrate is diffracted outside the dark zone by the second grating, reducing the total on-axis polarization leakage by multiple orders of magnitude. 
Because both beams are recombined, the phase pattern needs to produce a 360$^{\circ}$ dark zone, which requires more extreme patterns, increases the inner boundary of the dark zone, and generally reduces the PSF-core throughput.
However, the planet light is also recombined in the dark zone, so the throughput is a factor 2 higher. A common property is that all vAPP implementations is that they operate over 100\% bandwidth with extremely high efficiency ($>96$\%).\\
\textcolor{black}{The development of multiple types of vAPP coronagraphs coincided with progress in many other areas of the vAPP coronagraph, including vAPP design, the improvement of the focal-plane wavefront capabilities, and the manufacturing of gvAPPs for multiple high-contrast imaging instruments. 
In this paper we review this progress.
In section \ref{sec:design}, we detail the design considerations and we provide the first complete step-by-step description of the vAPP design process, including adding holograms, or wavelength selective behavior. 
We describe the impact of a standard gvAPP design on observation and data reduction in section \ref{sec:observe_data}, followed by an overview of the manufactured and planned vAPPs in section \ref{sec:world_of_vapp}. 
This includes their properties, details of the commissioning, performance, and for some, their first scientific results. 
In section \ref{sec:observing} we discuss the current status and provide suggestions for future upgrades. }

\section{Design of a gvAPP}
\label{sec:design}
The design of a gvAPP starts with the optimization of an APP phase pattern. 
\textcolor{black}{The goal of the optimization is to null the stellar PSF in the dark zone to by orders of magnitude with respect to the PSF core, e.g $10^{5}$}.
We define the contrast as this ratio of flux in the dark zone divided by the flux in the PSF core and it depends on the focal plane coordinates, see Fig. \ref{fig:APP_def}.
Light has to be diffracted from the stellar PSF core to null the dark zone, reducing the intrinsic Strehl ratio of the star.
We define the Strehl ratio as the summed flux of the non-coronagraphic stellar PSF core in an aperture divided by the summed flux of energy in the coronagraphic stellar PSF core using the same aperture. 
In simulation without noise this aperture has a diameter of one pixel, and for all other applications it is common to use a diameter of $\sim 1.4 \lambda/D$\cite{ruane2018}. 
Destructively interfering the star light in the dark zone results in constructive interference of the light on the opposite side of the PSF. 
A reduction of the stellar Strehl ratio is unwanted, as the Strehl ratio of the companion PSF is affected in the same way.
Therefore, the optimization simultaneously maximizes the Strehl ratio of the stellar PSF, while minimizing the flux inside the dark zone.
This optimization problem is highly non-linear in the complex phase exponential.
Consequently, first attempts at calculating APP patterns did not aim to find the optimal solution, their aim was to find solutions that are close.
These attempts used phase iteration techniques \cite{codona2004,codona2006} or a modified Gerchberg-Saxton (GS) algorithm \cite{kostinski2005} and were moderately successful.
They have shown that these methods can produce a dark zone of any shape with extreme contrasts ($<10^{-10}$) \cite{keller2016novel}, yet their Strehl ratios are low for small inner working angles.
Another set of solutions were generated by adapting the global optimization algorithms for shaped-pupil coronagraphs to include phase \cite{carlotti2013}.
This algorithm produces APPs with regions containing several discrete phases and sharp transitions in between these regions.
This algorithm was improved to yield \emph{globally} optimal solutions with unity amplitude across the pupil, smooth phase patterns for $180^{\circ}$ and D-shaped dark zones, and 0-$\pi$ solutions for $360^{\circ}$ dark zones \cite{por2017}.
As this last algorithm allows both phase and amplitude modulation in the pupil, this proves empirically that APPs are the globally optimal solutions for pupil-plane coronagraphs.
\subsection{Dark zone considerations}
\begin{figure}
    \centering
    \includegraphics[width=\linewidth]{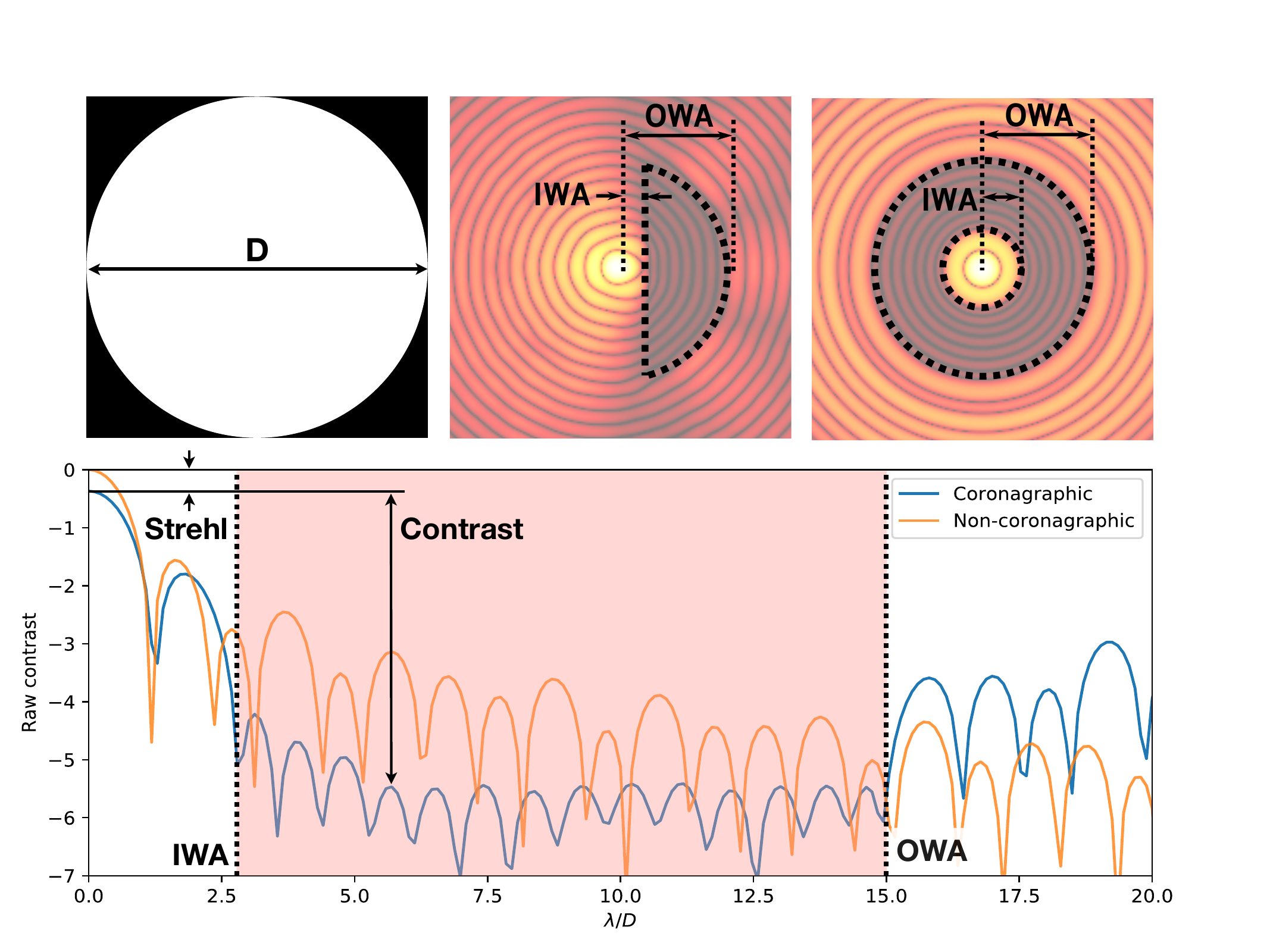}
    \caption{Definitions of the dark zone properties. \textit{Top left:} The pupil of the telescope. \textit{Top middle:} A D-shaped dark zone (log$_{10}$ scale).  \textit{Top right:} An annular dark zone (log$_{10}$ scale). \textit{Bottom:} PSF slice of a non-coronagraphic and coronagraphic design (annular). Adapted from Por 2017 \cite{por2017}. }
    \label{fig:APP_def}
\end{figure}

The globally optimal solutions are critical to ensure a high exoplanet yield.
However, it is not the full story.
As shown in \cite{por2017}, there is a trade-off between inner working angle (IWA), outer working angle (OWA), contrast, and Strehl. 
The definitions of the four properties can be found in Fig. \ref{fig:APP_def}.
For focal-plane coronagraphs, the IWA is defined as the smallest angular separation at which the total energy throughput of an off-axis source reaches 50\% of the maximum throughput \cite{ruane2018}.
For pupil-plane coronagraphs this can be adapted to the smallest angular separation where the planet throughput reaches 50\% of the maximum throughput in the region where the contrast reaches the target contrast.
Equivalently, the OWA is defined as the largest angular separation where the planet throughput still reaches 50\% of the maximum throughput in the region where the contrast reaches the target contrast.
A good choice of these parameters can be different for every planetary system, observing conditions, wavelength, telescope design and instrument performance. 
This becomes clear by using a metric that defines a planet detection given instrument limits. 
In Ruane et al. 2018 \cite{ruane2018}, the integration time $\Delta t$ that is required for a $1\sigma$ detection is given by
\begin{equation}
    \Delta t \propto \left[\frac{\eta_s + \sum_n a_n}{\eta_p^2-b^2/\epsilon^2}\right],
\label{eq:planetsnr}
\end{equation}
where $\eta_s$ fraction of available star light detected at the planet location, $\eta_p$ fraction of available planet light detected, and $\epsilon$ is the planet-to-star flux ratio. 
The factor $a_n$ represents noise factors like the background or detector noise such that $\sigma^2_n = a_n N_\star$, with $N_\star$ the total signal from star in photo-electrons. 
In the same way $\sigma_{\text{speck}} = bN_\star$, which represents the speckle noise.
In this framework, $\eta_p$ is directly proportional to the vAPP Strehl and the contrast is given by $\eta_s/\eta_p$. 
We do not minimize $\Delta t$ with vAPP optimization due to several complicating factors. 
However, we can use this equation to explore the impact of design choices.
Here we outline some of the considerations when choosing these parameters.\\
\emph{1. Strehl:} The Strehl ratio of a vAPP design impacts the planet throughput for the full field of view. 
In the case of the photon noise limit, Eq. \ref{eq:planetsnr} becomes $\Delta t = \eta_s/\eta_p^2$, demonstrating that Strehl greatly impacts the integration time that results in a 1$\sigma$ detection. 
The gvAPP already has the disadvantage of reducing the planet throughput by a factor of 2, so keeping the Strehl high is crucial for good performance.
The definition of a high Strehl is somewhat arbitrary given the impact of the shape of the telescope pupil, AO performance and presence of other noise sources. 
Therefore, we use the Strehl mostly for a trade-off between the other dark zone properties.\\
\emph{2. Contrast:} For ground-based high-contrast imaging systems, the intrinsic contrast of a coronagraph is almost never reached.
AO residuals and non-common path errors result in quasi-static speckles that limit the performance.
In addition, atmospheric jet streams create a wind-driven halo, as the temporal lag between the application of the wavefront correction and the evolving turbulence \cite{cantalloube2020}.
Designing a gvAPP with an intrinsic contrast much lower than the expected raw contrast does not yield the optimal performance.
This is clear from Eq. \ref{eq:planetsnr}, where planet is not detected in the presence of speckle noise if $\eta_p<b/\epsilon$ (negative integration time), independent of $\eta_s$.
Yet, removing bright PSF structures by creating a dark zone reduces speckle pinning, i.e. speckles that are spatially confined to secondary maxima in the diffraction limited PSF \cite{bloemhof2001}.
Taking into account speckle pinning, it can be argued that a design contrast lower than the expected AO-limited contrast does result in an improved performance, although that has not been studied in detail. 
For inner working angles $>2.0 \lambda/D$ and a central obscuration lower than 30\%, a design contrast of 10$^{-4}$ results in Strehl ratios higher than 70\% \cite{por2017}. 
This design contrast is already significantly lower than the raw contrast of extreme AO systems.
Another consideration is that constant contrast in the dark zone will yield the highest Strehl, although the changes in Strehl are minor for lower contrasts further out.
These considerations lead to the conclusion that a more optimal gvAPP design has a contrast that decreases gradually with the estimated AO residuals and becomes constant at a point where other noise factors take over, such as detector noise or background noise ($\sim 10^{-6}$). 
An example of such a design contrast is shown in Fig. \ref{fig:APP_def}.
In this case, the vAPP design contrast can be determined by AO simulations and verified with end-to-end simulations of the performance.\\
\emph{3. IWA:} All indirect detection methods for exoplanets (e.g. transit, radial-velocity, astrometry), show that there is a huge fraction of planets that is currently out of reach for direct imaging.
While these methods are biased to find these close-in planets, it shows that a small IWA is a critical to find new worlds with direct imaging. 
However, the Strehl ratio is also very dependent on the IWA.
For $10^{-6}$ contrast, an OWA of 8 $\lambda/D$, and a central obscuration of 10\% changing the IWA from 2.05 to 1.75 $\lambda/D$ reduces the Strehl from 60\% to 20\% \cite{por2017}. 
Furthermore, the gain of a smaller inner working angle is usually limited. 
Uncorrected low-order aberrations, both atmospheric or non-common path, reduce the performance at the smallest separations for ground-based telescopes. \\
\emph{4. OWA:} 
The outer working angle has limited impact on the Strehl of APP designs, while the dark zone area increases with the OWA squared. 
\textcolor{black}{The non-coronagraphic PSF contains only a small fraction of the energy at larger separations}, so increasing the contrast in this region results in a small decrease in Strehl as well.
In these regions the SNR of the exoplanets is not dominated by the contrast, however. 
Further out, techniques like ADI are more effective in removing speckles, the wind-driven halo and PSF structures. 
Thermal background in the near- and mid-infrared ($\lambda>2$ \textmu m) or detector noise start to dominate the SNR\textcolor{black}{, e.g. \cite{wagner2020}}.
It is therefore unnecessary for most applications to create a gvAPP with a OWA $>20\lambda/D$. \\
There is no general design that works for any telescope or wavelength range. 
All gvAPP designs are different because the optimal phase pattern depends on the telescope aperture and properties like the AO performance and background noise.
It is therefore advisable to perform a grid search to find the trade-offs between IWA, OWA, contrast at the IWA and the slope of the contrast as function of radius. 
An example grid could be the IWA between 1.8 and 2.3 $\lambda/D$ in steps of 0.1, the OWA between 12 and 20 $\lambda/D$ in steps of 1, the $\log($contrast$)$ between $10^{-3.5}$ and $10^{-5}$ in steps of 0.5, and a slope between 0.25 and 0.5 $\log($contrast$)$ per $\lambda/D$ in steps of 0.05.\\
Another consideration is the dark zone shape itself. 
A gvAPP with D-shaped dark zones provides phase solutions with higher Strehl ratios compared to 180 degree dark zones with identical inner working angles \textcolor{black}{\cite{por2017}}.
This difference in Strehl increases for smaller inner working angles ($<2.5 \lambda/D$), which is why the D-shape is used in most gvAPPs.
The impact of the D-shape on observing is discussed in section \ref{sec:D-shape}.
An interesting trade-off exists for planet detection at larger IWAs ($>3 \lambda/D$). Beyond this IWA threshold, the designs with an annular dark zone will have a Strehl ratio that can be competitive with D-shaped dark zones. 
While the Strehl ratio of designs with annular dark zones are still significantly lower, dgvAPPs with annular dark zones have twice the planet throughput compared to a gvAPP.
The origin of this factor two is that half of the planet light is imaged on the bright side of the coronagraphic PSF for a gvAPP.
The trade-off between D-shaped and annular dark zones is highly dependent on the telescope pupil, as a central obscuration size and spider thickness greatly impact the Strehl of annular dark zone designs.
We note that comparing Strehl ratios with a factor two correction factor for planet throughput and the same contrast levels does not necessarily select the best of the two.
There are other factors that are more difficult to add to this trade-off.
For example, a dgvAPP will not have wavelength smearing due to grating diffraction, companions will be inside the dark zone for all parallactic angles, post-processing is different, and a gvAPP requires a larger field of view.
Because factors like these are difficult to quantify, we did not attempt a general trade-off study.
So far, only one dgvAPP has been installed on the large binocular telescope (LBT).
The pupil of the LBT is favourable with a central obscuration ratio of $\sim11\%$ and thin spiders that were not included in the design.
Moreover, a vAPP with an annular dark zone was installed on the William Herschel Telescope (WHT), which was designed for an off-axis 1-m pupil without central obscuration and spiders.

\subsection{Optimization of the APP design}
\label{sec:vapp_optim}
The optimization of the APP starts with a pupil definition.
Existing HCI instruments usually have a pupil camera where the pupil can be measured or have their own mask to define the pupil.
Extreme caution is warranted when defining the pupil, as the maxima and minima of the phase pattern will be located near the pupil edges.
An error in pupil definition leads to a stark reduction in vAPP performance. 
For this reason we define an APP pupil, which is the undersized version of the true pupil to accommodate alignment errors, definition errors, and pupil movement. 
Drastic undersizing increases the IWA, negatively impacts Strehl with thicker spiders and a larger central obscuration ratio, and reduces throughput by removing effective telescope area.
Therefore, this balance results in an undersizing of the instrument pupil by 2-5\%, depending on the amplitude of the expected pupil movement and the alignment tolerances, and in coordination with the instrument team.
In addition, it can be beneficial to change the orientation of the dark zones depending on the pupil shape.
Spiders add narrow diffraction structures that locally enhance the PSF intensity. 
A D-shaped dark zone can, depending on the orientation with respect to the pupil, overlap with one or more spider diffraction structures.
Removing these structures leads to a decrease in Strehl, so if there is freedom to choose the dark zone orientation the straight edge should be oriented parallel to two spiders. 
Residual atmospheric dispersion, detector ghosting effects, the FOV of an IFS or the orientation of a slit or image slicer can limit this design freedom.\\
For a given set of APP parameters and a pupil definition, we use the optimization solver, Gurobi Optimizer \cite{gurobi}, to calculate the optimal solution. 
Computer memory and run time limitations limit the APP phase pattern to $100 \times 100$ pixels and a dark zone up to 14 $\lambda/D$.
\begin{figure}
    \centering
    \includegraphics[width=\linewidth]{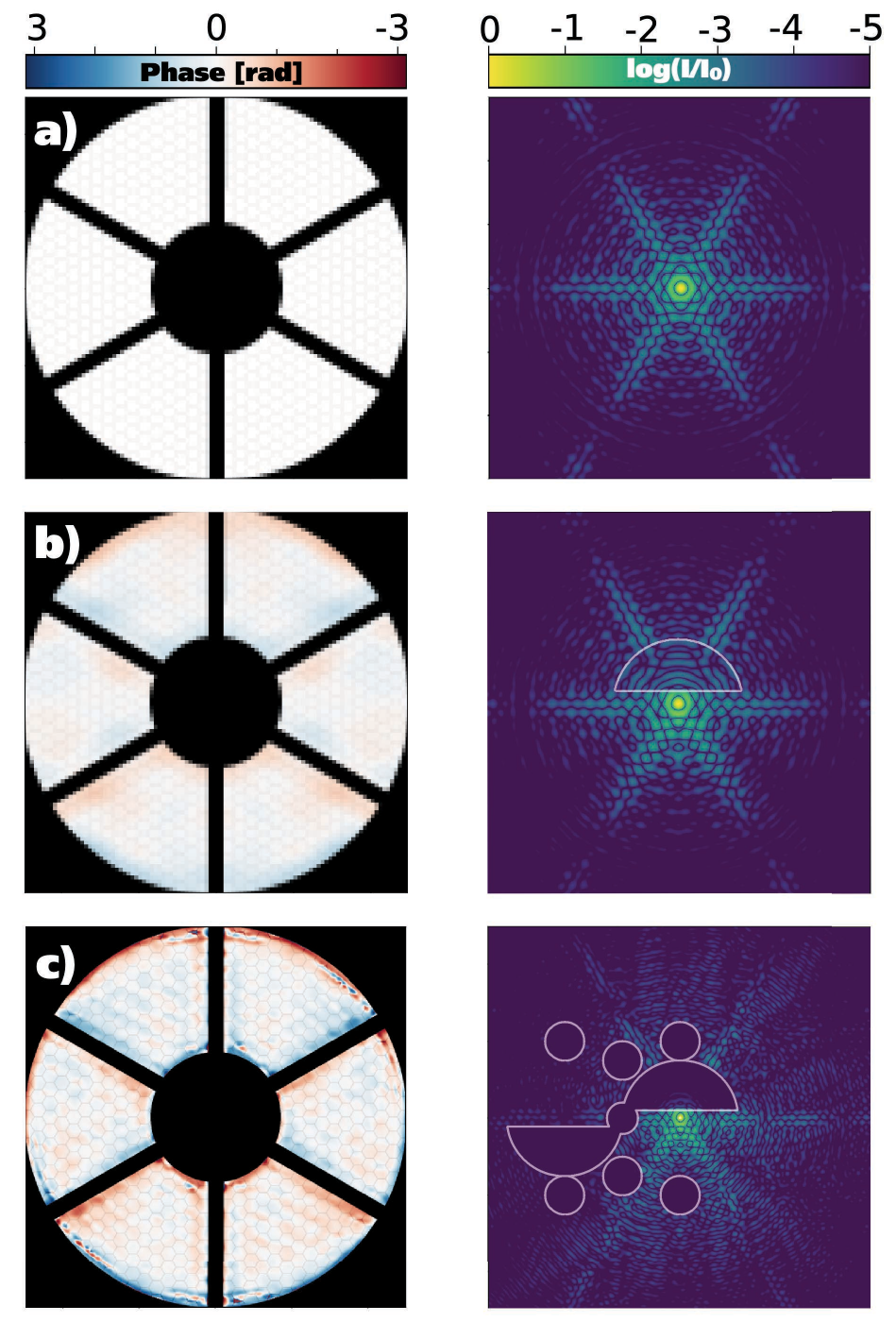}
    \caption{Optimization of an APP pattern, showing the phase pattern (left) and corresponding PSF (right). \textbf{a)} The low-resolution input pupil for the global optimizer and the corresponding PSF. \textbf{b)} Output of the global optimizer for a lower contrast ($10^{-3}$) and a smaller dark zone, indicated by the white lines. \textbf{c)} High-resolution APP design created using the Gerchberg-Saxton algorithm on the Fourier-scaled global optimal APP design. Additional dark zones have been added for the opposite coronagraphic PSF, the leakage PSF and additional holograms.}
    \label{fig:APP_design}
\end{figure}
For N pixels in the pupil plane, the optimization problem scales as N$^2$ in computer memory and N$^{3.5}$ in run time. 
These scaling laws also hold true for pixels in the focal plane for the OWA. 
Hence, it is not possible to calculate the APP design at the full resolution that is used in the direct-write system, which is on the order of a 1000$\times$1000 pixels.
We therefore adopt a two-stage approach for finding the vAPP phase pattern. 
First we find the optimal APP phase pattern for a downscaled version of the APP pupil.
If necessary we decrease the OWA to the 14 $\lambda/D$ limit set by the computer memory.
This downscaled version also does not contain additional dark zones for the coronagraphic PSF with opposite circular polarization or dark zones that minimize crosstalk with the leakage term or additional holograms. 
In section \ref{sec:holograms} the use of these additional dark zones and holograms is explained. 
Secondly, we upscale this low-resolution APP phase pattern design with Fourier upsampling \cite{soummer2007} and use this solution as a prior for the second stage.
This second stage adds all dark zones and corrects any errors made during the upsampling of the low-resolution phase pattern. 
We use a modified Gerchberg-Saxton (GS) algorithm \cite{gerchberg1972practical} for this stage. 
This algorithm does not guarantee an optimal solution like the global optimizer does. However, with the starting point already quite close the optimal solution, the GS algorithm is likely to converge before unacceptable Strehl losses occur.\\
An example for resulting phase patterns after the first and second stage are shown in Fig.~\ref{fig:APP_design}. In practice, we find that limiting the design contrast during the first stage of optimization often increases the Strehl of the final design. Typically, for a final design contrast $<10^{-4}$ the design contrast during the first stage should be between $10^{-2}$ and $10^{-3.5}$ to recover a new solution with the highest Strehl. Both optimization stages are implemented in \texttt{HCIPy} \cite{por2018} and an example of vAPP optimization using only GS can be found in its documentation\footnote{\url{https://docs.hcipy.org}}.
\subsection{Adding functionality with holograms}
\label{sec:holograms}
Enabled only by the accurate and high-resolution of the direct-write process and the liquid-crystal properties, it is possible to add capabilities to all types of vAPPs.
Here we will discuss two capabilities, focal-plane wavefront sensing (FPWFS) and reference spots for astrometry and photometry.
Both capabilities have solutions based on holograms.
We define holograms as PSF copies imaged off-axis, which can be biased with a wavefront aberration.
In essence, the vAPP PSFs are also holograms. 
Combining multiple holograms into a single phase screen is possible through multiplexing wavefronts into a single phase screen, as described in \cite{dong2012, wilby2017,doelman2018}.
A multiplexed phase screen directs light to multiple holograms, each with a different PSF.
These holograms can be added anywhere in the focal plane up to the Nyquist limit, with any bias, and with any amplitude mixing ratio. 
Due to circular-polarization splitting, adding a holographic copy at a certain position in the focal plane will necessarily also add a hologram with opposite bias at the location mirrored in the optical axis.
These properties of holograms make them very diverse and easily adapted for FPWFS and reference spots. 
A natural extension is to generate multiple coronagraphic PSFs per polarization state.
For example, it is possible to position two coronagraphic PSFs for one polarization such that they precisely overlap with the two coronagraphic PSFs of the opposite polarization state. 
The advantage is that the otherwise polarized coronagraphic PSFs now become unpolarized \cite{bos2020new}. 
This has benefits for including polarimetry and/or focal-plane wavefront sensing with the vAPP. 
One important downside is that multiplexing many holograms results in significant crosstalk. 
Crosstalk adds PSFs at locations that are the vector-addition of the individual gratings and multiples thereof. 
\subsubsection{Photometric and astrometric reference spots}
For accurate photometry of a companion the vAPP already has an advantage compared to focal-plane coronagraphs.
The coronagraphic PSFs themselves provide a photometric reference, given that the two PSFs are not saturated. 
Otherwise the leakage PSF can serve as photometric reference if it is not too affected by speckles.
However, it does have a different spectrum due to the wavelength dependence of the diffraction efficiency. 
A solution is an unbiased reference spot, which is a copy of the non-coronagraphic PSF. 
Multiplexing unbiased wavefronts with the vAPP results in reference PSFs that can be used both for photometry and astrometry. 
For other coronagraphs, such spots have been generated using the deformable mirror or static phase screen.
A full summary of these efforts can be found in Bos (2020) \cite{bos2020} and references therein.
Moreover, the vector speckle grid that is proposed in Bos (2020) \cite{bos2020} produces speckles that are effectively incoherent with the underlying halo, which greatly improve the photometric and astrometric accuracy.
By multiplexing the vAPP with two holograms with opposite phase modulation on the opposite polarization states, this vector speckle grid is easily implemented in the vAPP coronagraph.
Astrometry with a vector speckle grid could provide astrometric solutions with a precision of $<0.01 \lambda/D$ \cite{bos2020}, however this has not yet been tested on-sky.
\subsubsection{Focal-plane wavefront sensing}
Adding a bias wavefront to holograms changes their sensitivity to the incoming wavefront.
Holographic modal focal-plane wavefront sensing is possible because of these variations \cite{keller2016novel}.
FPWFS is critical to remove non-common path aberrations (NCPA), aberrations from optics after the AO system, which are therefore unseen by the AO wavefront sensor. 
An example of a focal-plane wavefront sensor is the coronagraphic modal wavefront sensor (cMWS) \cite{wilby2017, haffert2018}, which provides simultaneous coronagraphic imaging and focal-plane wavefront sensing with the science point-spread function.
The cMWS creates multiple holograms, each biased with a different wavefront mode drawn from a suitable basis set.
The normalized difference between the PSF copies with opposite circular-polarization state responds linearly to the corresponding aberration mode present in the input wavefront.
The main advantage of this wavefront sensor is that the monitoring the Strehl ratios of the holograms and the wavefront reconstruction are straightforward and can be done at high speeds and with the actual science camera, independent from the dark hole(s).
Moreover, no modulation of the wavefront is required.
Multiplexing holograms removes light from the coronagraphic PSF, and for efficient wavefront sensing each mode takes away $\sim 1\%$ of Strehl, on top of the already lower vAPP Strehl.
The trade-off between modal coverage, sensitivity and Strehl is difficult to optimize, as it is a priori unclear how much the selected wavefront modes contribute to speckles. 
The cMWS has successfully been tested with the Leiden exoplanet instrument (LEXI) \cite{haffert2018}, and in lab demonstrations of Magellan/MagAO-X \cite{miller2019}. \\
Another method for FPWFS is an adaptation of phase diversity (PD) \cite{Gonsalves1982}.
Where classical PD requires an in-focus and out-of-focus image, phase diversity holograms are biased with defocus of opposite amplitude, similar to the cMWS with a larger bias.
This larger bias results in an increase in modes that can be reconstructed, while decreasing the average intensity of the hologram.
So compared to the cMWS, a larger fraction of the incoming light is diffracted to the PD holograms, e.g. 10\% instead of 1\%.
This FPWFS method was implemented in the Subaru/SCExAO\cite{doelman2017}, Magellan/MagAO-X\cite{miller2018development} and HiCIBaS instruments. 
\begin{figure}
    \centering
    \includegraphics[width=0.8\linewidth]{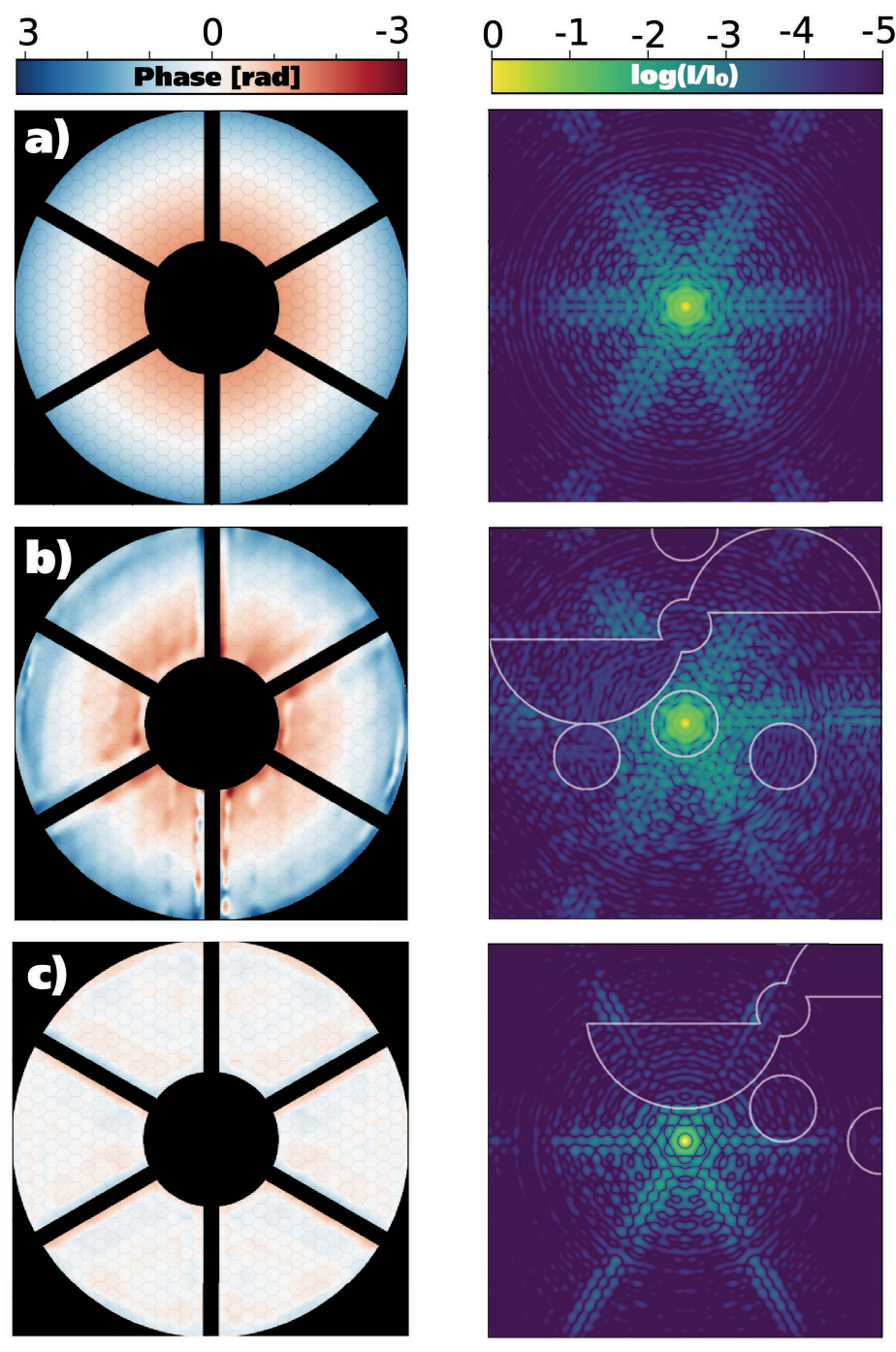}
    \caption{Updates of the holographic phase patterns to minimize crosstalk and light scattering into the dark zones. $\textbf{a)}$ An unaltered phase diversity hologram with a defocus of 0.87 radian RMS. \textbf{b)} The updated PD hologram with suppressed diffraction structures in all dark zones, calculated using the Gerchberg-Saxton (GS) algorithm. \textbf{c)} Same as b), only without defocus to create a photometric or astrometric reference hologram. Each hologram has displaced dark zones. This displacement corresponds to the relative location of the hologram with respect to the dark zones of the coronagraphic PSF after multiplexing the phase patterns.}
    \label{fig:hologram_design}
\end{figure}
\begin{figure}
    \centering
    \includegraphics[width=\linewidth]{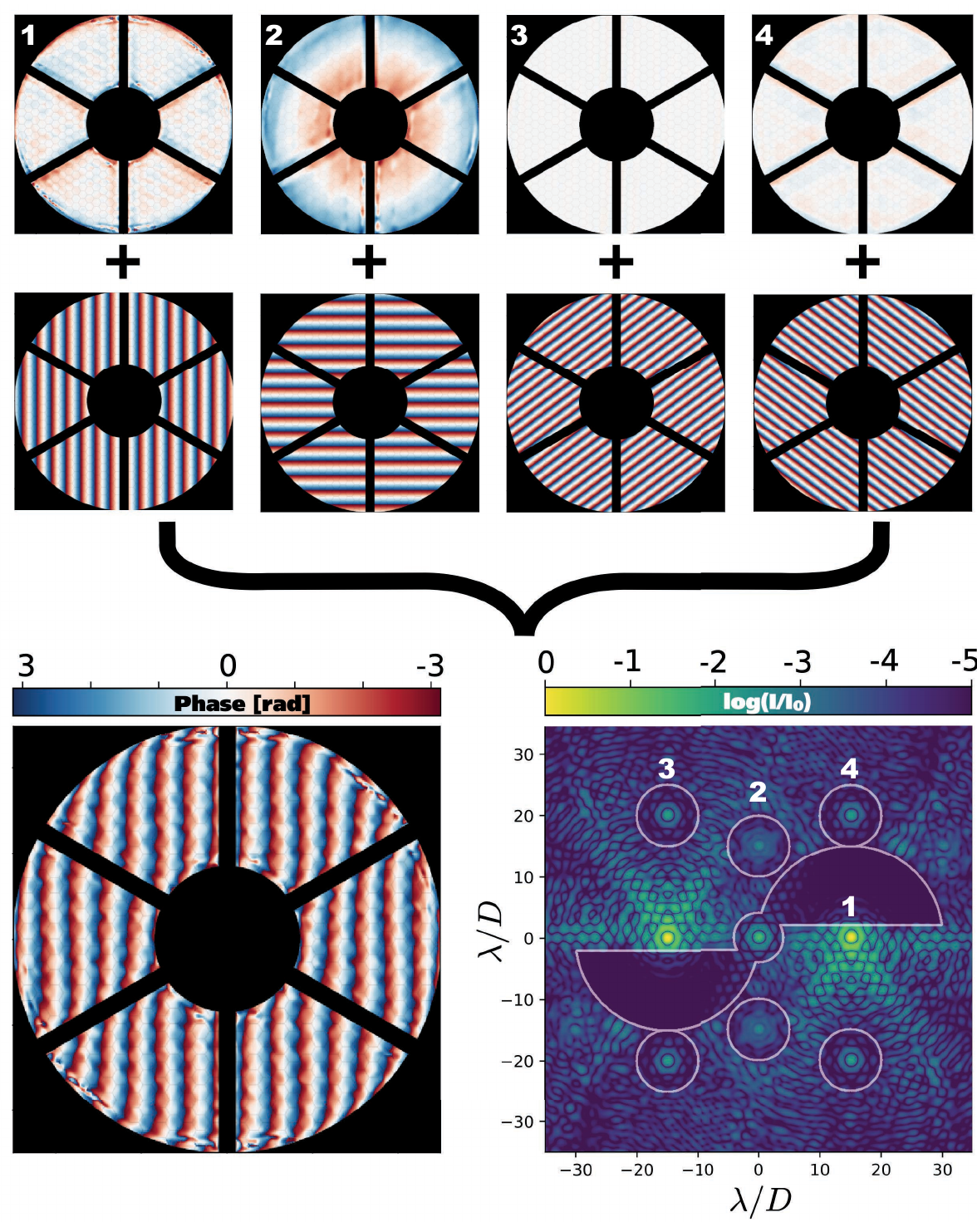}
    \caption{Creation of a gvAPP phase pattern by multiplexing the APP design (1) with the three holograms. Hologram 2 is defocused and could be used for focal-plane wavefront sensing. Hologram 3 and 4 are unaberrated and are located in a square grid at the grating frequency for astrometric referencing.  Additional iterations with GS have cleared the dark zone of second order effects caused by multiplexing. The PSF is calculated assuming unpolarized light and a polarization leakage of 1\%.}
    \label{fig:multiplexed_vapp}
\end{figure}
A different implementation of PD is described in Por (2016) \cite{por2016}, directly reconstructing the focal-plane electric field in the dark zone.
Coronagraphic PSF copies are created with each a unique probe in the dark zone, making use of the pairwise nature of the holograms \cite{keller2016novel}.
The unaltered science PSF is still available, providing the same advantages as the cMWS, but adding the capability to sense all modes corresponding to the dark zone.\\
Interestingly, the coronagraphic PSFs themselves can also be designed to encode wavefront information in the bright field by including a pupil-plane amplitude asymmetry. 
The added advantage of the vAPP is that there are two coronagraphic PSFs, with opposite bright fields that cover all of the spatial frequencies. 
This reduces the number of modes that can otherwise not be measured by one coronagraphic PSF \cite{sun2019efficient}. We refer to such vAPP designs as Asymmetric Pupil vAPPs (APvAPPs), and was successfully implemented by Bos et al. (2019) \cite{bos2019}.
They show that if the pupil is asymmetric, both odd and even modes can be recovered from the coronagraphic PSFs only.
Earlier versions of an asymmetric pupil wavefront sensor \cite{martinache2013asymmetric} could not be combined with a coronagraph.
A non-linear model-based wavefront sensing algorithm was proposed by \cite{bos2019} and is limited, due to computational considerations, to the first $\sim100$ Zernike modes, yet it showed nearly an order of magnitude improvement in contrast between 2 and 4 $\lambda/D$ with the internal source, and on-sky the gain was a factor of 2.
Spatial linear dark field control (LDFC) \cite{miller2017,miller2019} is also a very powerful combination with the APvAPP. 
LDFC empirically derives a linear relationship between DM actuation and focal-plane intensity response in the bright field, allowing it to control a larger number of modes. 
As LDFC can only drive back the wavefront to a reference state, it requires an initial calibration by non-linear model-based wavefront sensing algorithm, making both methods complimentary.
LDFC uses the linear response to wavefront aberrations in the bright side of the vAPP PSF to reconstruct the wavefront.
For Magellan/MagAO-X, simulations have shown that LDFC is capable of correcting the full dark zone using 400 modes \cite{miller2019}.
Recently, LDFC in combination with an APvAPP was installed and tested using Subaru/SCExAO \cite{miller2020spatial}.
An on-sky demonstration followed soon after \cite{bos2021submittedfirst}.
By multiplexing two coronagraphic PSFs per polarization state, each of these PSFs designed with a different pupil-plane amplitude asymmetry, it is also possible to design APvAPPs that can measure both pupil-plane phase and amplitude modes simultaneously \cite{bos2020new}.  
The possibility of adding photometric, astrometric, and wavefront sensing capabilities to the vAPP make it a unique and diverse coronagraph. 
Exploring these techniques for a more integrated approach to building high-contrast imaging systems shall lead to a better performance at the systems level.
\subsubsection{Implementation into vAPP designs}
Adding holograms by multiplexing phase patterns does not result in an optimal combination. 
Crosstalk between modes and the vAPP PSF will decrease the contrast in the vAPP coronagraphic dark zones.
Moreover, the crosstalk will change the response of the holograms. 
Minimizing this crosstalk is implemented by changing the vAPP PSF, the modes and optimizing after multiplexing.
First, the coronagraphic PSF can be suppressed at the locations of the holograms, such that interference is minimized.
Additional dark zones can be added during the upscaling of the APP design.
Second, the modes can be adapted to also have a reduced intensity at the location of the APP dark zone. 
These dark zones can be added using the GS algorithm, as is shown in Fig. \ref{fig:hologram_design}.
All modified holograms, including the coronagraphic vAPP hologram, can then be combined by multiplexing.
Third, the crosstalk in the vAPP dark zones caused by the multiplexing can be removed by applying an additional round of GS. 
The combined result is shown in Fig. \ref{fig:multiplexed_vapp}.
Every time GS is applied, it has a minor impact on the Strehl of the coronagraphic PSF. 
The final cleaning of the dark zone could impact the performance of holographic wavefront sensors, though we estimate that this effect is minimal, as no significant deviation from expected performance was found in \cite{miller2019}

\subsection{Additional design choices}
So far we have focused on the design of the vAPP phase pattern. 
However, there are additional choices regarding the vAPP optic that are independent of the phase pattern.
We will highlight two design choices that impact the manufacturing complexity and the wavelength dependent performance of the vAPP.
   \begin{figure} 
   \includegraphics[width=\linewidth]{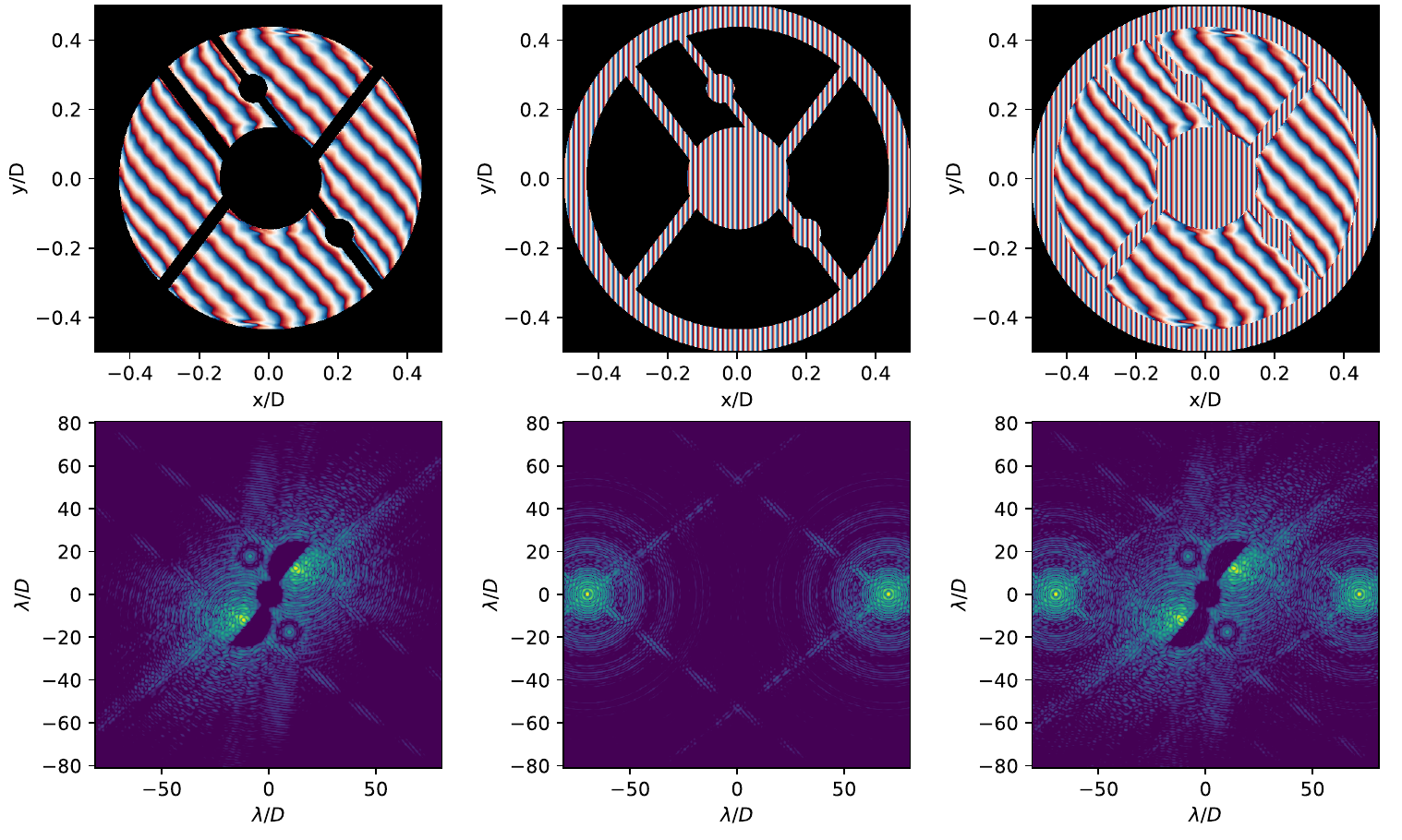}
   \caption[example] 
   { \label{fig:gratingmask}
   \textit{Top left:} Phase of a grating-vAPP for an undersized pupil of the Subaru Telescope. An amplitude mask for the pupil is shown in black. \textit{Bottom left:} Logarithmic plot of the PSF of the grating-vAPP, showing two coronagraphic PSFs for opposite handedness of circularly polarized light. The grating-vAPP was simulated to be a perfect half-wave retarder, so no leakage is present. \textit{Center:} The grating mask (top) and resulting PSF (bottom) that will be used as an amplitude mask. The grating has 90 periods across the diameter of the pupil. \textit{Right:} The phase and PSF of the grating-vAPP combined with the grating mask. The incoming light has a pupil of radius 0.45 (black) and the light is separated in a grating mask component and a grating-vAPP component. A small misalignment of the pupil does not affect the coronagraphic PSF, relaxing alignment tolerances. Adapted from Doelman et al. 2017 \cite{doelman2017}.}
   \end{figure}
\subsubsection{The grating mask}
\begin{figure} 
   \includegraphics[width=\linewidth]{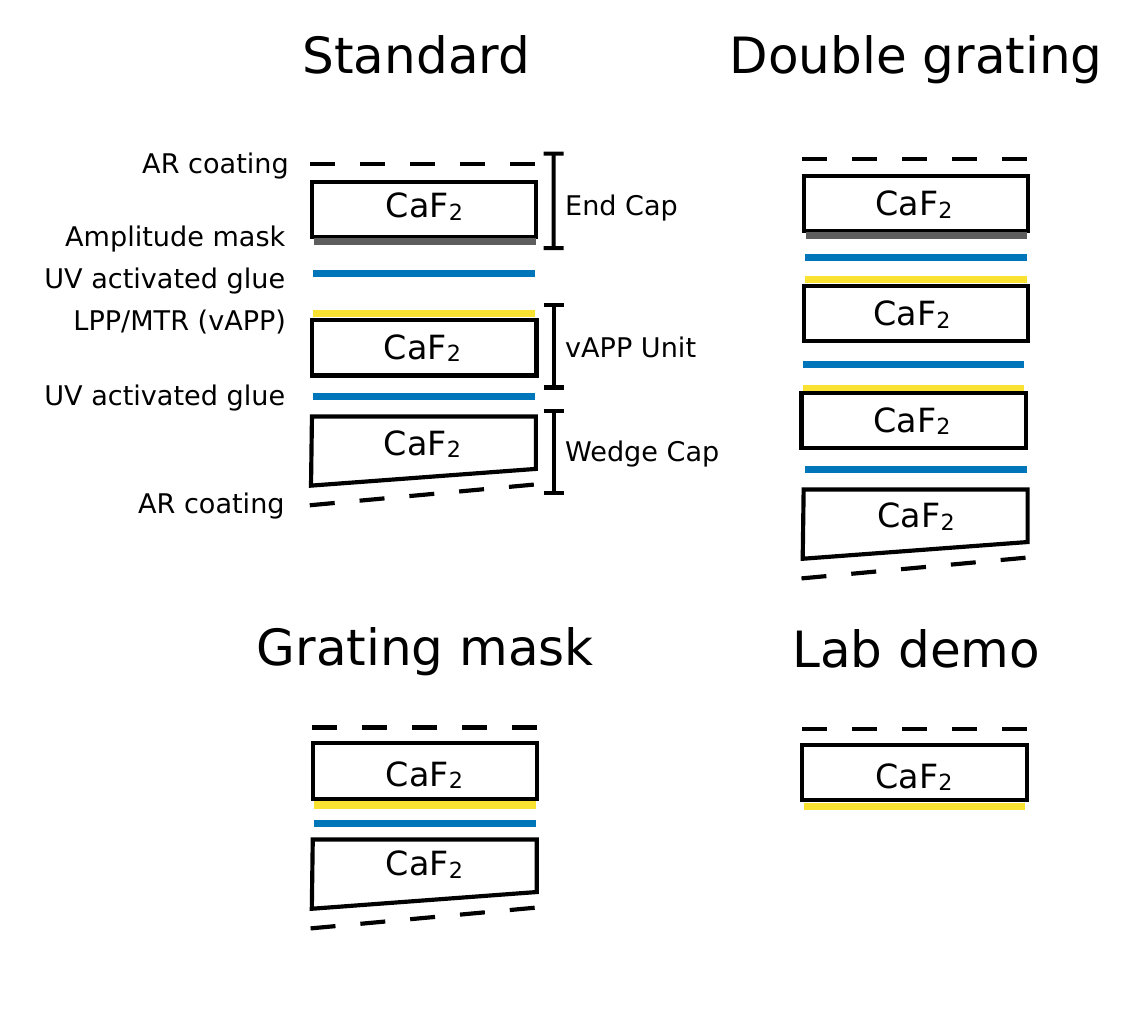}
   \caption{Various substrate designs for vAPP manufacturing. }    
  \label{fig:vapp_assembly}
\end{figure}

As discussed in section \ref{sec:vapp_optim}, the pupil definition is critical for the performance of the vAPP. 
During manufacturing it is therefore standard to add a physical binary amplitude mask to the assembly that is aligned with sub-micron accuracy with the phase pattern. 
These amplitude masks and their alignment add significant complexity and cost to the vAPP optic. 
Especially for lab demonstration projects that often require multiple phase patterns, it is not cost effective to use these amplitude masks.
Separate metal laser-cut masks can provide a cheap solution at the cost of reduced alignment accuracy.
We note that it is not necessary to absorb or reflect the light at the pupil plane.
A different solution is to diffract the light outside of the field of view of the coronagraphic PSFs.
Diffracting the light that is not falling onto the coronagraphic phase pattern to outside the FOV requires a patterned high-frequency polarization grating outside of the pupil, e.g. beyond the outer diameter, inside the spiders, and within the central obscuration.
This grating should have a different orientation and frequency than the grating of the gvAPP.
A simple spatial filter in an intermediate focal plane can remove the light outside the pupil. 
We call this concept the ``grating mask''. 
This concept is described in detail in Doelman et al. 2017 \cite{doelman2017}, and a reworded version is added here for clarity.
An example of the gvAPP with a grating mask is shown in Fig. \ref{fig:gratingmask}. 
Critical for the workings of a PG as amplitude mask is that the tilt is applied \emph{locally}, as the phase is accumulated inside the liquid-crystal layer \cite{escuti2016}.
This is unlike a regular grating where the interference between the \emph{full} grating results in the diffraction orders. 
Adding a mask only outside the pupil does not significantly alter the coronagraphic PSF. 
It does, however, change the leakage term from the PSF of the undersized vAPP pupil to the PSF of the telescope or instrument pupil. \\
How the grating mask reduces the complexity of the vAPP optic can be seen in Fig. \ref{fig:vapp_assembly}. 
With a grating mask the first substrate is no longer necessary, removing a glue layer as well. 
Moreover, the lab demo vAPPs can be manufactured without any additional substrates. 
While their phase patterns and pupils can be extremely complex, only homogeneous illumination is necessary for them to work as expected. 
We note that an end cap would make them more durable and less easy to damage. 
\begin{figure} 
   \includegraphics[width=\linewidth, trim={0cm 5.5cm 0cm 5.5cm},clip]{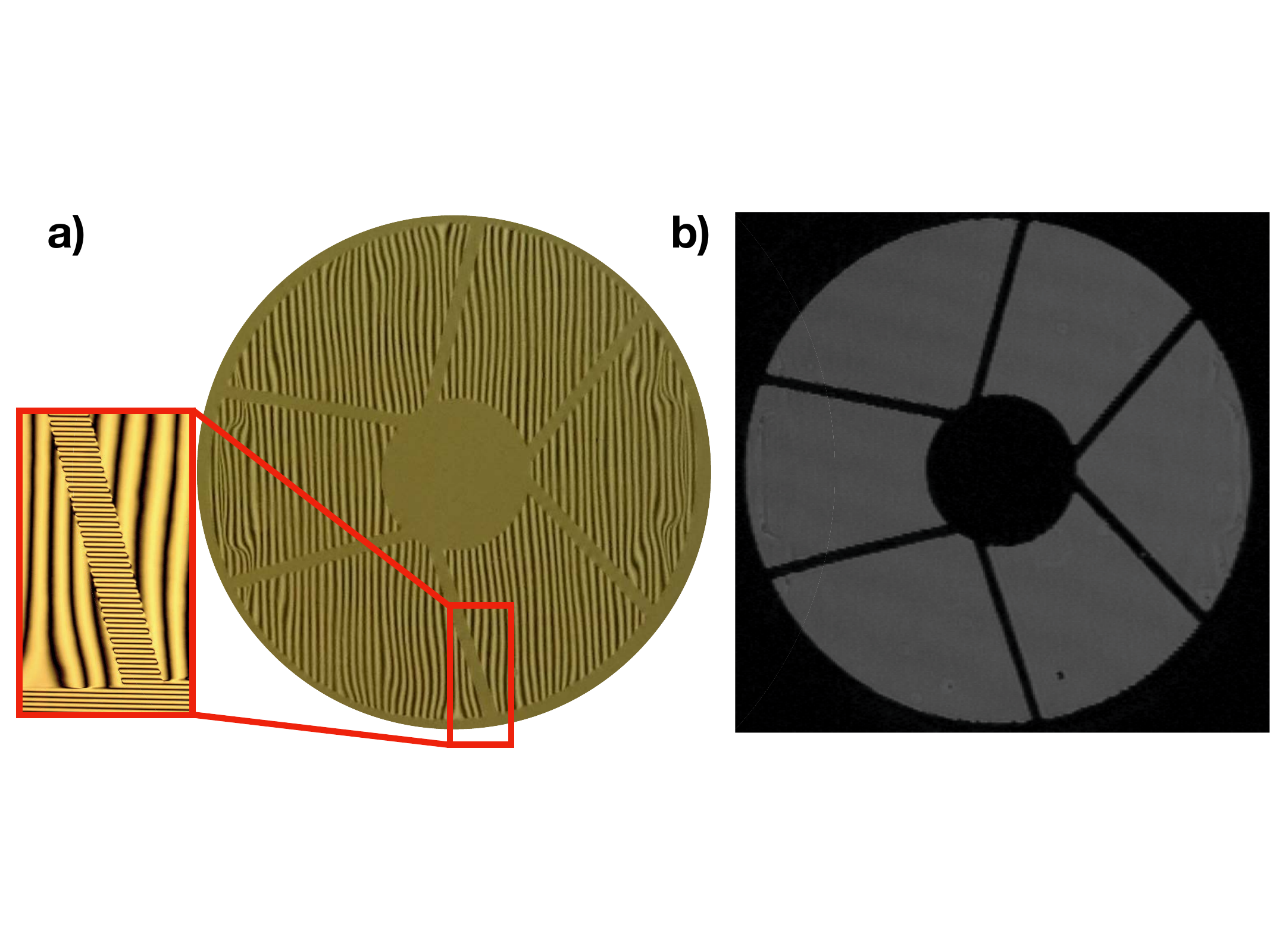}
   \caption[example] 
   { \label{fig:imagegm}
   \textbf{a)} Photograph of the vAPP with a simplified RST pupil between crossed polarizers, manufactured for testing the grating mask. The zoomed in microscope image shows the high-frequency grating used for the grating mask that is not captured by the normal camera. \textbf{b)} Reimaged pupil with a field stop inserted in the focal plane. The field stop removes the diffracted light from outside the pupil defined by the grating mask. Even spiders with a width of a few times the grating period are removed extremely well. }
\end{figure}
For lab testing a 1 cm grating mask was produced with 8 pixels of 8.75 micron per period.
The pupil is a simplified version of the Nancy Grace Roman Space Telescope and was chosen for the thin spiders and complex pupil design.
The full mask between crossed polarizers is shown on the left in Fig. \ref{fig:imagegm} in addition to a microscopic image which shows a spider with the grating pattern.
The width of the spider is a few times the grating period, yet this does not affect the performance of the grating mask.
The right panel of Fig. \ref{fig:imagegm} shows the reimaged pupil when a field stop is added to the focal plane. 
The light outside the simplified RST pupil, which is diffracted by the grating, is blocked and does not appear in the reimaged pupil. \\
When applied in a real high-contrast imaging system, two things need to be considered. 
First, one has to be careful where the diffracted light from outside the pupil is directed to.
The diffracted light can scatter back onto the detector by reflections inside the instrument if not blocked.
Because the direction of the PG is a free design parameter, the reflections can be directed to non-reflective parts with prior knowledge of the instrument layout. 
Second, a few percent of the light that would be normally blocked by an amplitude mask would end up on the science camera due to polarization leakage.
This polarization leakage is always non-zero for a deviation in retardance from half-wave.
The influence of the polarization leakage will be small when the coronagraphic PSFs are have enough spatial separation from the leakage.
This separation is also a free parameter and is set by the frequency of the grating of the coronagraphic phase pattern.
In addition, the pupil defined by the grating mask will be undersized a few \% compared to the telescope pupil, so the amount of light that is diffracted by the grating mask is limited.\\
Overall, the grating mask is a cheap alternative for amplitude masks and alleviates the problems that come with aligning an amplitude mask with the phase pattern. 

\subsubsection{Wavelength selective multi-twist retarders}
Another option for the vAPP involves the tuning of the retardance in a non-standard way. 
For most applications the retardance is tuned to be close to half-wave over the full bandwidth of an instrument \cite{komanduri2013, otten2017} to maximize the efficiency.
However, by changing the liquid-crystal recipe one can, in principle, make the retardance follow any continuous curve as function of wavelength  \cite{hornburg2014multiband}.
We note that this is possible because the fast-axis orientation and retardance of the devices are decoupled. 
\begin{figure} 
\begin{center}
\begin{tabular}{c} 
\includegraphics[width = \linewidth]{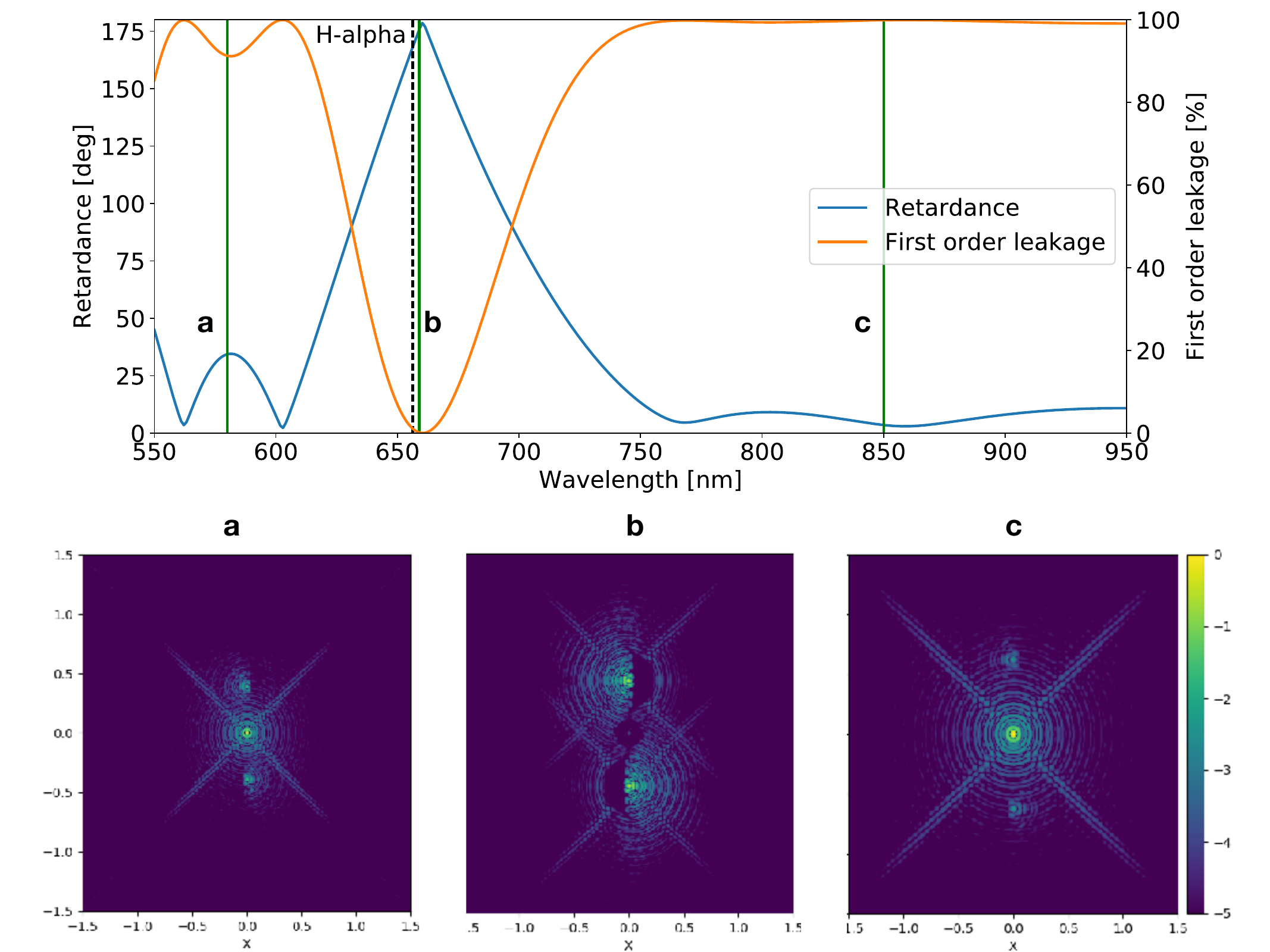}
\end{tabular}
\end{center}
\caption[example] 
{
\textit{Top:} Simulated retardance (blue) and first order leakage (orange) as a function of wavelength for a 3-layered MTR. The total thickness is 20.7 micron and it was optimized to have 0 retardance everywhere except for a 10 nm band around H$\alpha$ wavelength (656.28 nm), where the retardance is close to 180 degrees. \textit{Bottom:} The simulated PSFs of a grating-vAPP with the retardance profile shown above. The green lines in the top figure correspond to the wavelengths of the bottom panel. At wavelengths shorter than H$\alpha$ the leakage term dominates and almost no coronagraphic PSF can be seen (a). This is the same for wavelengths larger than 750 nm (c). Around H$\alpha$ the leakage term disappears and the gvAPP operates as normal (b).\label{fig:wavelength-selective}
}
\end{figure}
One specific application of a liquid-crystal recipe is the ``wavelength-selective coronagraph'' \cite{doelman2017}.
Here we include a short summary of this work.
A wavelength-selective coronagraph has rapidly changing efficiencies, where the retardance switches between intervals of 180 degrees as function of wavelength.
A retardance of zero or a full wave results in a zero efficiency, such that the light traveling through the coronagraph does not acquire geometric phase.
The coronagraph is effectively switched off at those wavelengths, while it is switched on in the range where the retardance is close to half-wave. 
This technology has been demonstrated for RGB lenses \cite{hornburg2019highly}. 
An example for the vAPP coronagraph is shown in Fig. \ref{fig:wavelength-selective}. 
The retardance is close to 180 degrees for H$\alpha$ and close to zero for smaller than 600 nm and larger than 750 nm. 
The light outside the H$\alpha$ spectral band can be used by a wavefront sensor, enabling wavefront sensing close to the science bandwidth with optimal efficiency.
In addition, the wavefront sensor can be placed after the coronagraph to minimize non-common path aberrations.
This requires a simple focal plane mask that separates the central term from the coronagraphic PSFs, aided by some spectral filters to minimize spectral cross-talk. 

\section{Observing and data reduction}
\label{sec:observe_data}
The gvAPP can be described as a simple coronagraph, as it is only a single pupil-plane optic that suppresses the diffraction halo.
However, there are two differences between other coronagraphs that make both observing and data reduction much less straightforward, i.e. the D-shaped dark zone and the grating. 
For the D-shape dark zone of a gvAPP the dark zones do not cover 360 degrees, which takes away a larger fraction for closer-in orbits caused by sky-rotation. 
This has to taken into account for planning observations and in the data reduction process. 
In addition, the grating can smear the planet PSF over multiple pixels in the grating direction, reducing signal-to-noise and the effectiveness of observing techniques like angular differential imaging (ADI) \cite{marois2006angular}.
\subsection{Influence of the D-shaped dark zone}
\label{sec:D-shape}
\begin{figure}
    \centering
    \includegraphics[width = \linewidth]{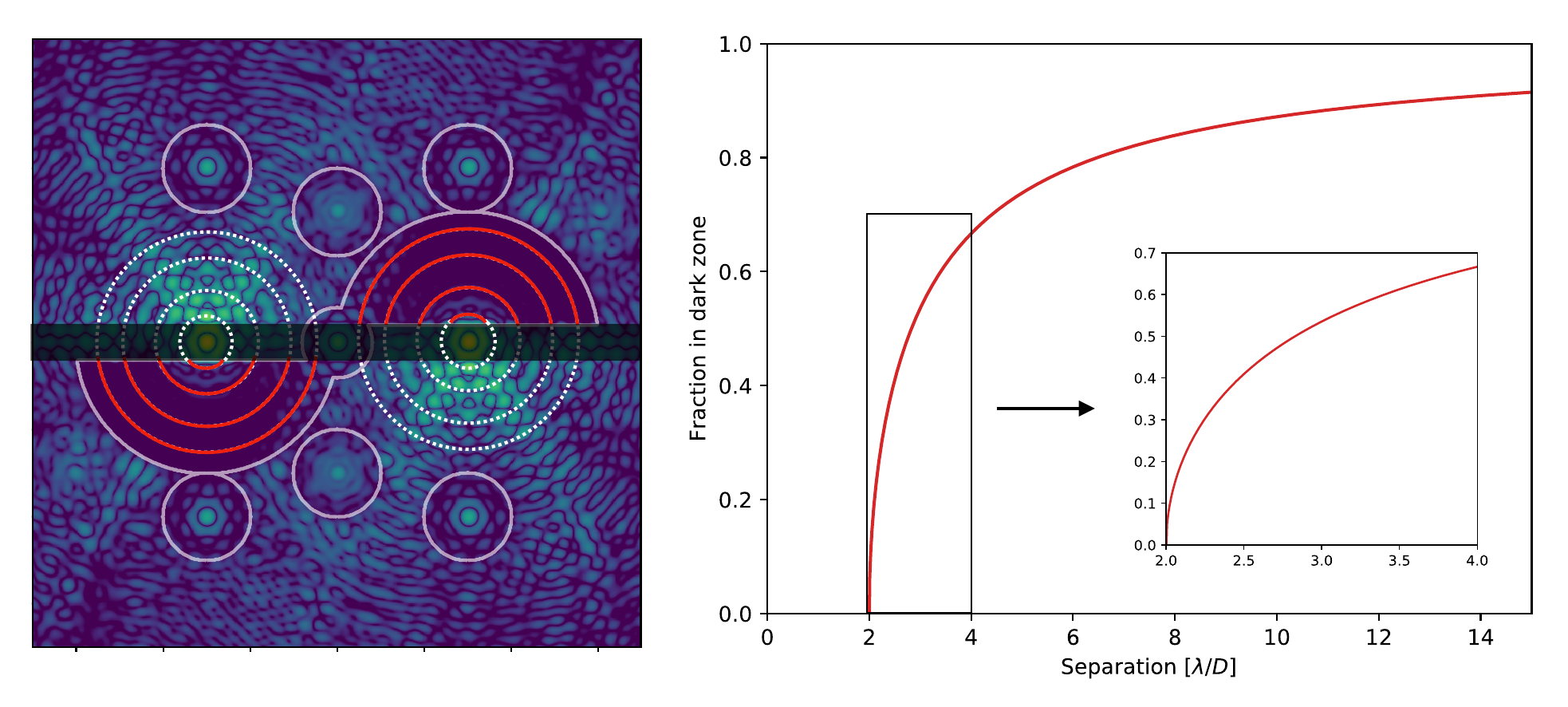}
    \caption{Planets rotating in the dark zone in pupil-stabilized mode are not always in the dark zone. \textit{Left:} Fraction of a circle (white) that is in the dark zone (red). \textit{Right:} The same fraction as function of separation for an inner working angle of 2$\lambda/D$. }
    \label{fig:dztime}
\end{figure}

The D-shaped dark zone provides phase solutions that have much higher Strehl ratios for similar smaller inner working angles than a 180 degree dark zone, while providing a similar total area.
We note that the term inner working angle is somewhat misleading here.
It is correct that the D-shaped and 180 degree dark zone designs can detect a planet down to the same separation, yet the \emph{amount of time} it is in the dark zone for pupil tracking mode is very different.
For a 180 degree dark zone the coverage is a full 360 degrees, while a bar is missing for the D-shaped dark zone, which takes away a larger fraction for closer-in orbits.
The fraction of a circle, $F_{DH}$, with radius $r$ that is inside a dark zone with an inner working angle IWA is given by
\begin{equation}
F_{DH} = \frac{\pi - 2\arcsin{(\text{IWA}/r)}}{\pi}.
\end{equation}
This fraction is also shown in Fig. \ref{fig:dztime} for a typical gvAPP design.
On-sky the rotation speed can change dramatically near Zenith, and the position in the dark zone depends on the local sidereal time at the observatory.
Therefore, observations with the gvAPP should be carefully planned if a known target is observed.
Planning does not help with detecting unknown objects where certain areas around the star will have more noise due to the sky-rotation placing objects in the bar between dark zones.\\
For known objects the blind spots created by the incomplete coverage of the D-shaped dark zones can be predicted. 
To this end, we created a preliminary version of an observation preparation tool in \texttt{python}, which can be found on \texttt{Github} \footnote{\url{https://github.com/18alex96/vAPP_preparation}}. 
The goal of this tool is to predict the locations of known companions and provide information on when it is in the dark zone to help determine observation strategies. 
The current version has an internal library that contains the information of a few operational gvAPPs, such that a simple query returns the predicted locations of companions in a PSF image. 
The code needs the instrument name, target name, the companion position angle and separation, and the observation times.
Wavelength scaling has not been implemented yet.
The core of the observation tool is based on \texttt{Astropy} \cite{price2018astropy}, which is used to retrieve object information and coordinate transformations. 
\subsection{Influence of wavelength smearing}
The grating of the gvAPP is an elegant solution to separate the polarization leakage from the coronagraphic PSFs. 
It does, however, represent a challenge for the data reduction. 
For narrow-band data it displaces the two coronagraphic PSFs which need to be recombined after centering them separately.
This is similar for data taken with an integral field spectrograph (IFS), except the instrument provides simultaneous multi-wavelength observations.
For each wavelength, the stellar PSF and companion PSF will have shifted and also needs to be recombined.
This can be characterized, especially if there are no optical distortions in the IFS such that the displacement from the grating is linear. 
Wavelength smearing is likely to reduce the astrometric accuracy in addition to make object spectra more sensitive to the lenslet flat.
For larger bandwidths the wavelength smearing introduces more problems.
The light of the planet is smeared over many pixels, resulting in an increase in camera noise and a larger sensitivity to speckle noise.
In addition, the planet PSF rotates around the stellar PSF for every wavelength, which is a different location on the detector for every wavelength.
In the derotated rest frame of the stellar PSF the orientation of the elongation changes in time, such that ADI techniques are less effective.
Collapsing the PSF in the grating direction before derotation might solve this problem, but this does not remove the sensitivity to speckle and camera noise.\\
The classical vAPP, which separates the two polarization states with a Wollaston prism, would not have these issues caused by the grating \cite{snik2012}.
However, they would be limited by polarization leakage at the smallest separations for large bandwidths.
A double-grating vAPP is also not affected by the grating, it is not limited by polarization leakage, and does have full 360 degree FOV at the cost of a larger inner working angle.
Combining these two techniques creates a solution for the wavelength smearing and large inner working angle.
\textcolor{black}{This combination then requires a polarizing beam splitter, like a Wollaston prism, to separate the two polarization states of a double-grating vAPP with D-shaped dark zones \cite{bos2018fully}}.
\subsection{Data reduction}
\begin{figure}
    \centering
    \includegraphics[width = \linewidth]{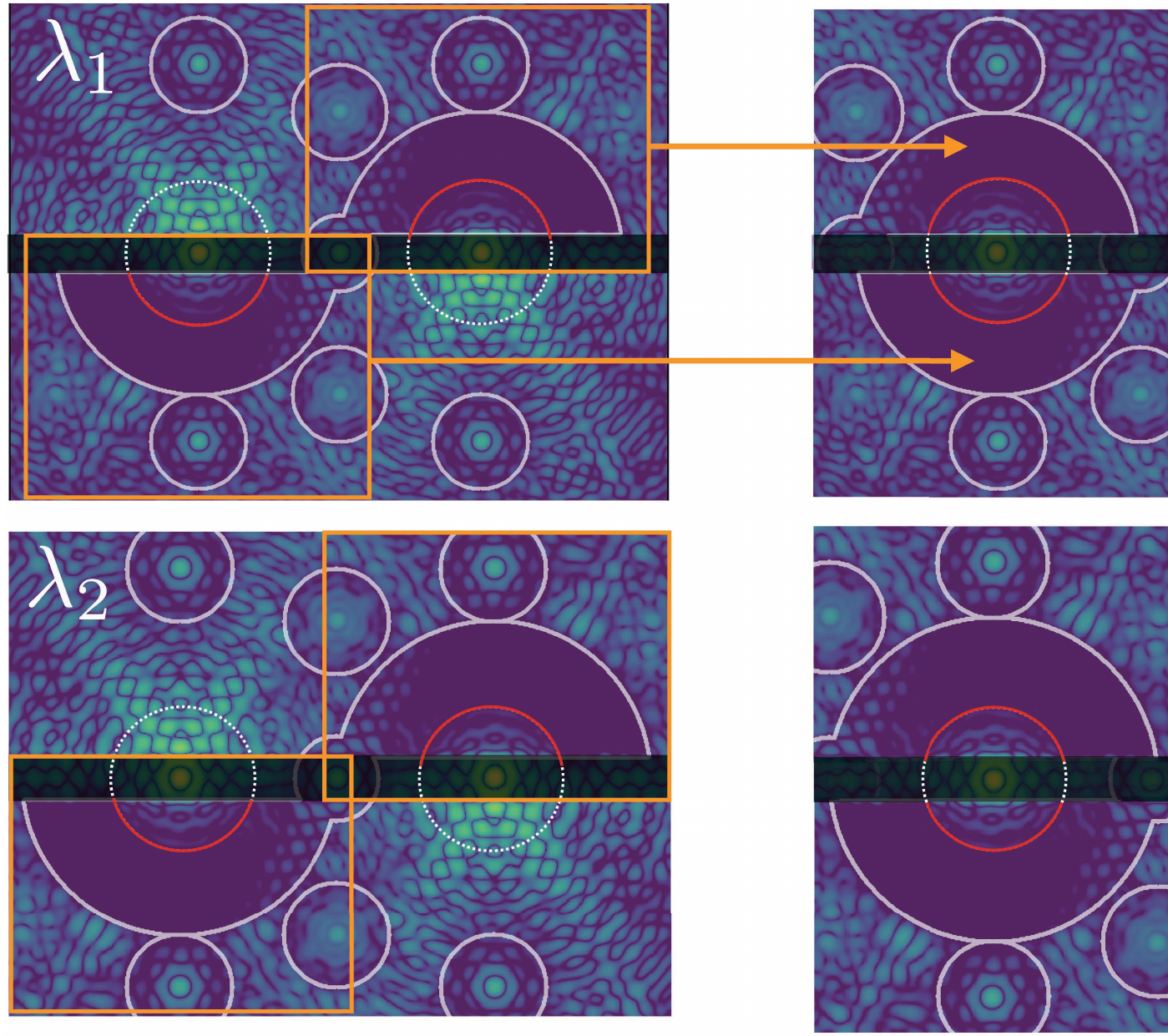}
    \caption{Method of recombining both coronagraphic PSFs such that the data can be processed using conventional data reduction techniques. }
    \label{fig:vAPP_data_combination}
\end{figure}
From the complications of data reduction caused by the D-shape of the dark zone and the grating, it is clear that gvAPP data can not simply be reduced with commonly-used data reduction pipelines like the Vortex Image Processing package  \cite{gonzalez2017vip} or PynPoint \cite{amara2012pynpoint,stolker2019pynpoint}. 
These pipelines include different data reduction algorithms that have been developed for data taken without coronagraphs or with focal-plane coronagraphs. 
In both situations the stellar PSF is not separated and the search space is 360 degrees. 
However, with a relatively simple adaptation it is possible reshape gvAPP data in a way that these pipelines can be used.
We create a window for each coronagraphic PSF and combine them into a single new data frame that is centered on both PSF cores.
An example is shown in Fig. \ref{fig:vAPP_data_combination}.
This method also works for IFS data if the windows change location with wavelength, but does not solve the wavelength smearing for large bandwidths. 
In addition, it does not make use of the fact that both coronagraphic PSFs are aberrated by the same wavefront aberration.
The rotate subtract algorithm, suggested by \cite{otten2017}, provides an additional contrast gain by using this symmetry.
An improved version of this algorithm does not simply subtract the average PSF but uses \textcolor{black}{principal} component analysis \cite{wold1987principal} to create a PSF library using one coronagraphic PSF and optimally removing these components from the other PSF.
\textcolor{black}{Another technique that is likely well suited for gvAPP data is the Temporal Reference Analysis for Exoplanets (TRAP) algorithm \cite{samland2019high}}.
TRAP creates a data driven model of the temporal behavior of the systematics using reference pixels, which can also be the pixels at the other coronagraphic PSF. 
\textcolor{black}{Moreover, this algorithm assumes a PSF model that can include wavelength smearing, which is a significant problem for the gvAPP. 
Therefore, TRAP presents an exciting opportunity to improve the gvAPP data reduction.}
\section{The world of vAPP}
\label{sec:world_of_vapp}
Over the last decade the vAPP has been installed in 7 instruments on 6 telescopes, and will be installed in two instruments on the Extremely Large Telescope (ELT).
The success of the vAPP can be explained by two of its properties: simplicity and adaptability.
Each vAPP is unique and fully optimized for the system it is implemented in, both in wavelength coverage as phase design, while only a single optic needs to be installed.
The vAPPs cover almost an order of magnitude in bandwidth, i.e. 0.55-5 $\mu$m, provide multiple ways of focal-plane wavefront sensing, and provide photometric/astrometric capabilities.
First, we will provide an overview of the different properties and locations of the vAPPs.
Then, we will shortly describe the unique capabilities of each vAPP.
\subsection{The vAPP: a global view}
\begin{figure}
    \centering
    \includegraphics[width = \linewidth]{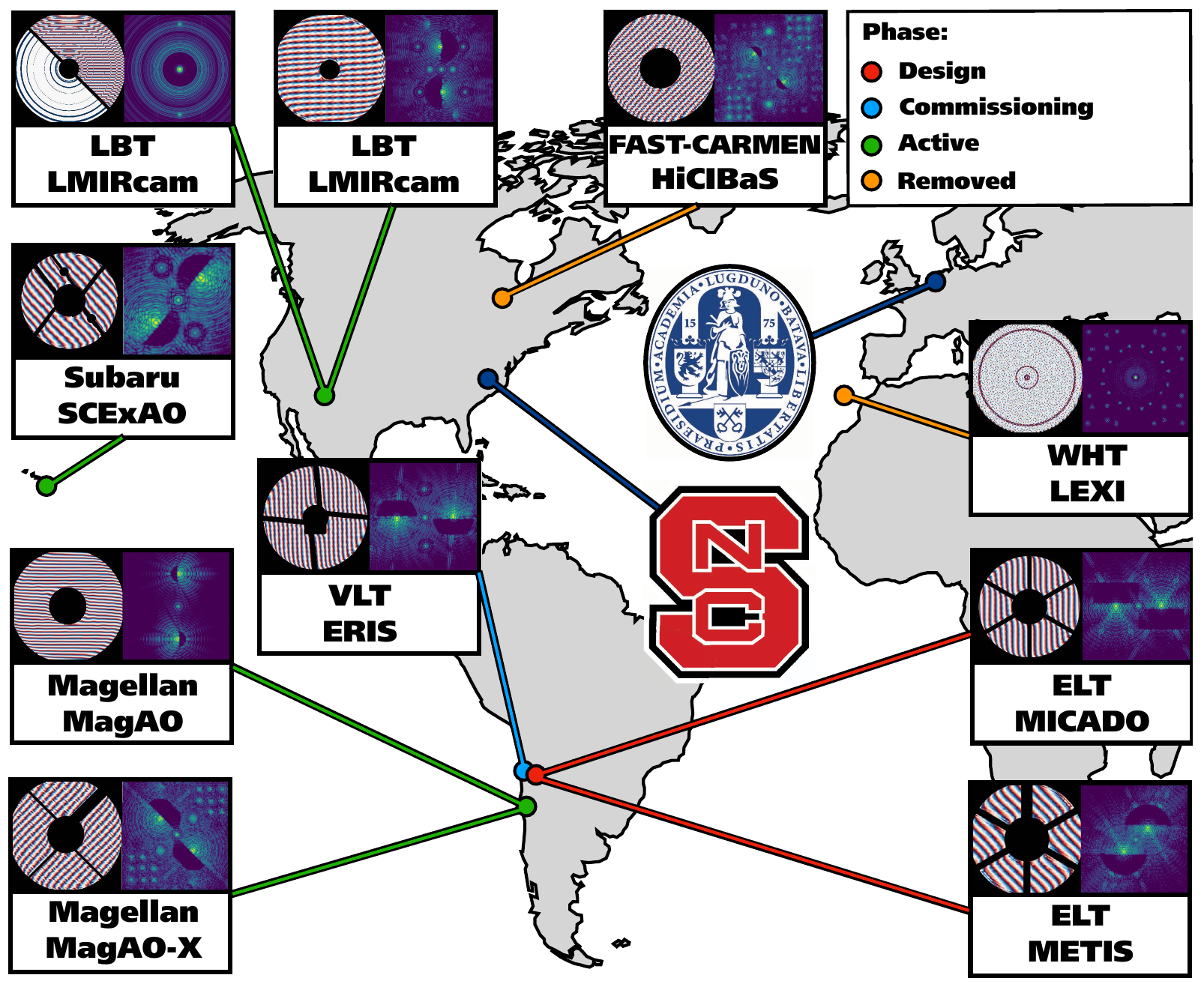}
    \caption{Overview and status of the vAPP coronagraphs around the world.}
    \label{fig:world_of_vapp}
\end{figure}

\begin{figure}
    \centering
    \includegraphics[width = \linewidth]{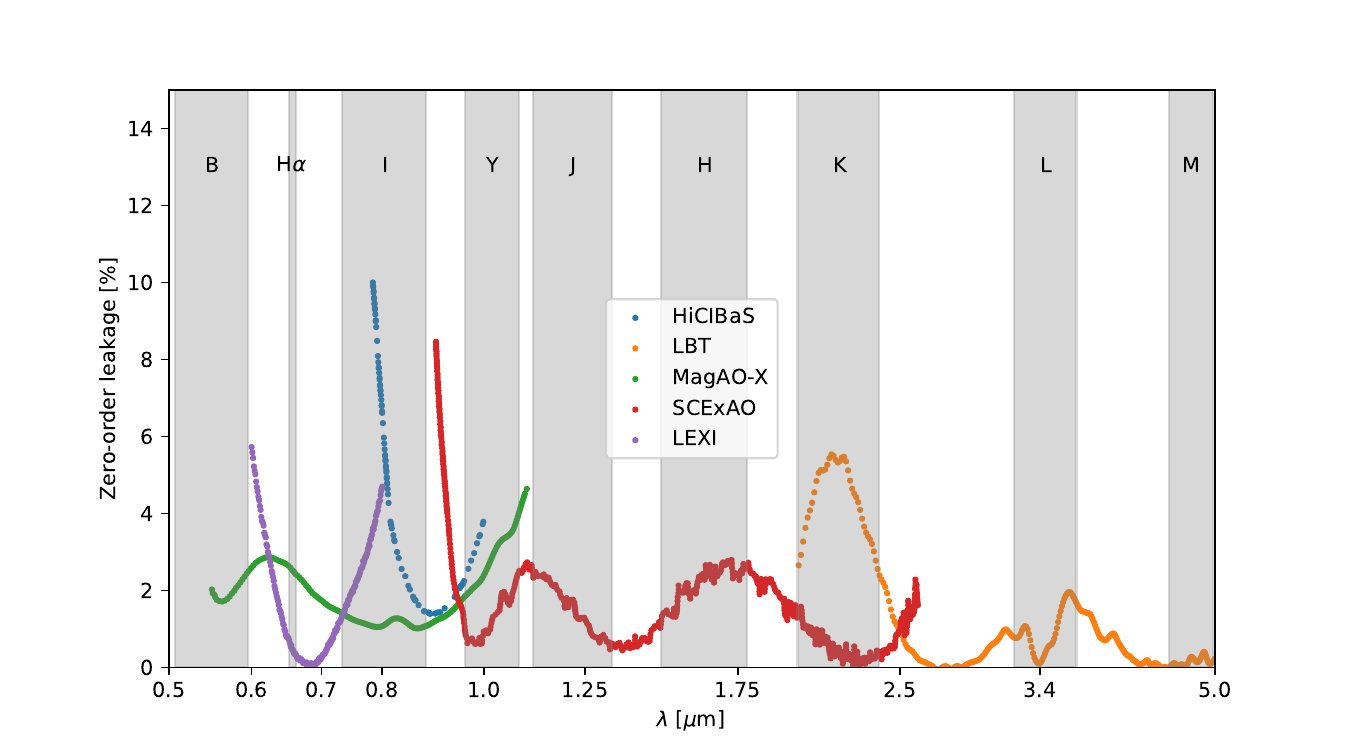}
    \caption{Zero-order leakage as a function of wavelength for different liquid-crystal recipes, colored by instrument.}
    \label{fig:LC_lin}
\end{figure}

High-contrast imaging systems are now part of most 8m-class telescopes. The vAPP has benefited from the enormous growth in the number of high-contrast imaging systems.
A global view of the vAPP coronagraphs is shown in Fig. \ref{fig:world_of_vapp}.
The vAPPs are operational in both the Northern and Southern hemisphere.
For each vAPP, we show the phase pattern and the simulated PSF.
In addition, we display the project status, indicated by the color of the marker on the map.
As of 2020, the vAPP has been tested in two former instruments, is operational in four instruments and will be installed on three more instruments.
A summary of most properties of all vAPPs are presented in Table \ref{tab:vapp_prop}.\\
\clearpage
\begin{sidewaystable}[h!]
    \centering
    \caption{Properties of the vAPP coronagraphs. FPWFS techniques: 1 = AP-WFS \cite{bos2019}, 2 = cMWS \cite{keller2016novel,wilby2017,miller2019}, 3 = PD \cite{Gonsalves1982,miller2018development}. }
    \begin{tabular}{l|l|c|c|c|c|c|c|c|c|c}
         Telescope & Instrument & Type & Inst. & $\lambda$ & MTR & Contrast & IWA  & OWA  & Strehl & FPWFS \\
         &&&&[$\mu$m]&&&[$\lambda/D$]&[$\lambda/D$]&[\%]&\\
         \hline
         \hline
         Magellan & MagAO/Clio & gvAPP 2x180& 2015 & 2-5 & 3 &$10^{-5}$ & 2.0 & 7 & 39.8 & - \\
         WHT & LEXI & vAPP 360 & 2017 & 0.6-0.8 & 1 &$10^{-4}$& 2.5 & 6 & 57.2 & 2 \\
         Subaru & SCExAO/& gvAPP 2x180 & 2018 & 1-2.5&3&$10^{-5}$&2.1 & 11 & 49.2 & 1+3 \\
         &CHARIS&&&&&&&&\\
         FC & HiCIBaS & gvAPP 2x180& 2018 & 0.83-0.88 & 1 &$10^{-6}$& 2.1 & 8.5 & 28.8 & 2+3 \\
         Magellan & MagAO-X & gvAPP 2x180& 2018 & 0.55-0.9 &3&$10^{-5}$ & 2.1 & 15 & 40.7 & 1+2+3 \\
         LBT & LMIRcam & gvAPP 2x180& 2018 & 2-5 &3& $10^{-5}$ & 1.8 & 15 & 52.6 & 3 \\
         LBT & LMIRcam/ & dgvAPP 360 & 2018 & 2-5 &3& $10^{-5}$ & 2.7 & 15 & 42.0 & - \\
         &ALES&&&&&&&&\\
         VLT & ERIS & gvAPP 2x180 & 2022 & 2-5 &3&$10^{-5}$& 2.2 & 15 & 50.9 & 1 \\
         ELT & METIS & gvAPP 2x180 & 2028 & 2.9-5.3&3 &$10^{-5}$& 2.5 & 20 & 63.8 & 1 \\
         ELT & MICADO& gvAPP 2x180& 2026 & 1-2.5&3&$10^{-5}$& 2.6 & 20 & 68.8 & - \\
    \end{tabular}
    \label{tab:vapp_prop}
\end{sidewaystable}
\clearpage
Another important property of the vAPP is the polarization leakage as function of wavelength.
The design of the liquid-crystal recipes aims to minimize polarization leakage over the full bandwidth and are given in Table \ref{tab:vapp_prop}.
For small bandwidths ($<15$\%), the liquid-crystal design can be kept simple with only single layer with zero twist. 
Larger bandwidths require multiple layers, each with different thickness and twist, to form a multi-twist retarder (MTR). 
In general, if the bandwidth is between $15-40$\%, a double-layered recipe, i.e. a 2TR, can have a polarization leakage $<3\%$ for the full bandwidth.      
Similar performance can be reached with a 3TR for bandwidths up to 100\%.\\
When a specific design is chosen, the manufacturing process is tuned to generate a liquid-crystal film with a measured retardance and polarization leakage that mimics the theoretical as closely as possible.
The manufactured test films are polarization gratings (PGs), such that it is easy to measure the zero-order leakage using transmission measurements.
When a recipe is perfected, the vAPP is manufactured using this recipe.
Measuring the zero-order leakage of the final optic can be cumbersome with the small angular separations between the main beams and the leakage, caused by the large grating periods of a gvAPP.
Therefore, the leakage measurement of a representative PG is used.
These polarization leakages of five different recipes, representative of 5 different instruments and wavelength ranges, is shown in Fig. \ref{fig:LC_lin}.
All presented recipes have $<6\%$ leakage over their required bandwidth.
\begin{figure}
    \centering
    \includegraphics[width = \linewidth]{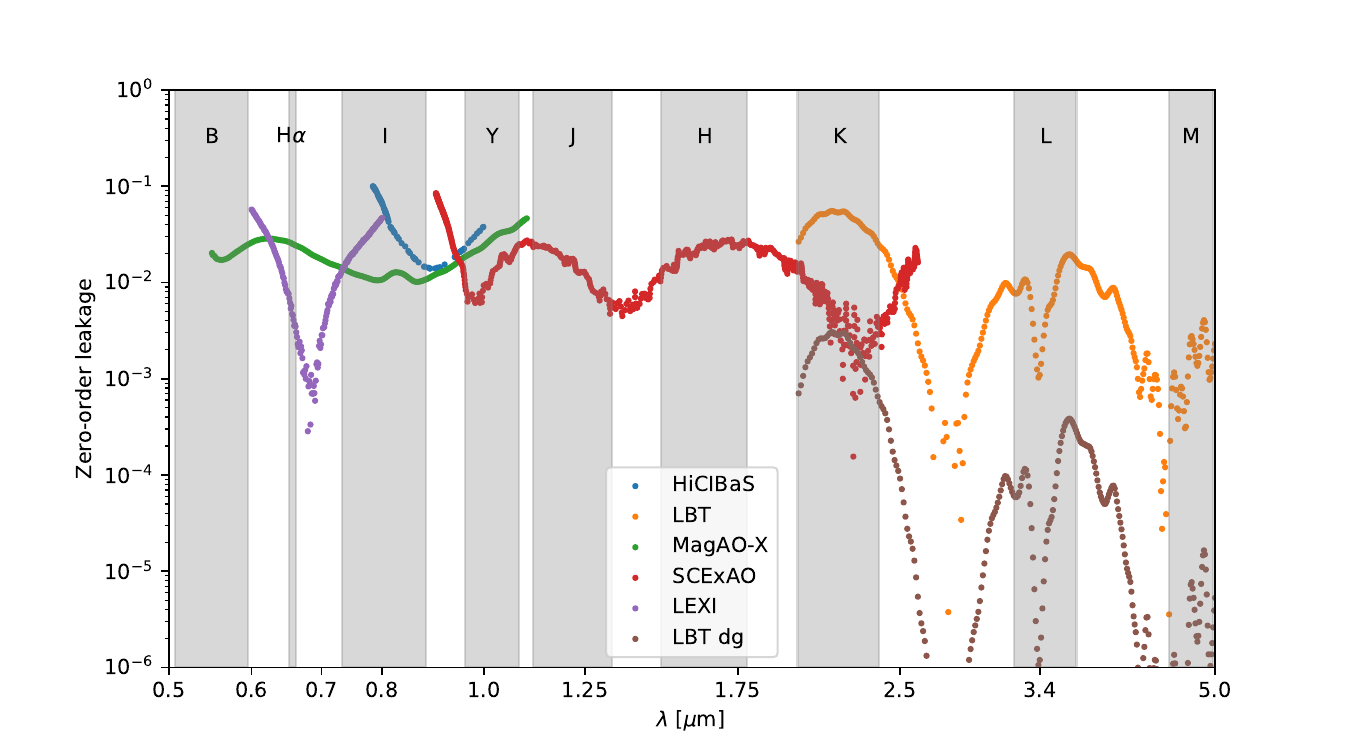}
    \caption{Same as Fig. \ref{fig:LC_lin}, but on log-scale and an additional estimate of the LBT double-grating vAPP.}
    \label{fig:LC_log}
\end{figure}
Together, they span the almost the full visible and near-infrared (0.55-5 $\mu$m).
Interestingly, we currently cover this range with three standardized recipes that have good performance in their respective bandwidths.
The same data is presented in log-scale in Fig. \ref{fig:LC_log}. 
We added the estimated performance of the double-grating vAPP that is installed in the LBT, by multiplying the single-grating curve with itself. 
We note that no direct measurement of its performance exists.\\
While the leakage measurements are representative, the actual vAPPs can have different performances. 
Lab conditions during manufacturing, like temperature and humidity, can generate differences between liquid-crystal films for identical recipes.
Characterizing the actual polarization leakage of a vAPP can then be done after installation or in a high-contrast imaging setup.
Moreover, all vAPPs with identical wavelength coverage will have different polarization leakages as function of wavelength. 
These differences can be up to 2\% for a specific wavelength, but the impact on the average is minimal.
Most likely, it is merely a shift of the maxima and minima as function of wavelength.
\subsection{The vAPP: individual properties}
A large diversity exists between vAPPs, in designs, design goals, additional functionality, and the instruments that use them.
We will provide a summary of design trade-offs, unique properties, historical details, notable results, and references to more detailed papers of each vAPP.
\subsubsection{Magellan/MagAO/Clio}
The first operational broadband gvAPP was installed in 2015 in the MagAO/Clio instrument \cite{close2010,morzinski2014} at the 6.5-m Magellan Clay telescope.
The gvAPP has a dark zone between 2-7 $\lambda$/D and operates between 2 to 5 micron, with a polarization leakage less than 3\% over this bandwidth.  
A detailed description of the design, manufacturing and on-sky performance can be found in Otten et al. (2017) \cite{otten2017}. 
Being the first gvAPP to be commissioned, it was a technology demonstrator with the additional goal to use it as a science-grade component. 
To demonstrate the performance, Otten et al. observed a bright A-type star with an $L'$-band magnitude between 0 and 1.
Characterization of the leakage term and the PSF showed that the gvAPP was performing as expected.
A new method of data reduction, subtracting the opposite rotated coronagraphic PSF, yielded an improvement in contrast of 1.46 magnitudes, i.e. a factor 3.8, at 3.5$\lambda$/D. 
The performance of the gvAPP for this observation was better than the best reported performance in literature of coronagraphs at other systems at this small angle.
We note that the method of calculating the noise is necessarily different between the contrast curves reported in literature, as noted in Jensen-Clem et al. (2017) \cite{jensen2017}.
The method of noise calculation in Otten et al. (2017) \cite{otten2017} does not take into account azimuthally correlated speckle noise caused by non-common path aberrations, pointing jitter, and thermal variations.
For a gvAPP it is especially difficult to properly estimate the speckle noise at the smallest separations due to the limited amount of photometric apertures that fit in the D-shaped dark zones. 
A better comparison between this gvAPP and other coronagraphs would be to use the same detection limit methods that are commonly used, e.g. fake planet injection.\\
So far, the scientific yield of the MagAO/Clio gvAPP is limited, with only a marginal detection of a brown dwarf companion of HR 2562 \cite{Sutlieff2019,sutlieff2021}.
This can be explained by the total awarded observing time ($<$10 nights), which is insignificant in comparison to the SPHERE and GPI surveys ($>$300 nights) that detected on the order of 10 sub-stellar companions \cite{nielsen2019gemini,vigan2020sphere}.
Moreover, observing with a gvAPP can be challenging, especially when using broadband filters, see section \ref{sec:observe_data}.\\
\subsubsection{WHT/LEXI}
The Leiden Exoplanet Instrument (LEXI) was a visiting instrument at the William Herschel Telescope (WHT) \cite{haffert2016,haffert2018}. 
The goal of the LEXI instrument was to develop and test new technology for new high-contrast imaging techniques, e.g. the cMWS and high-contrast, high-dispersion integral field spectroscopy \cite{wilby2016,wilby2017,haffert2018}.
The instrument changed between different visits, changing the pupil (4m on-axis, 1m off-axis), the wavefront sensor, imaging modes, spectrographs and vAPP designs.
The first observation run in 2016 employed a spatial light modulator with an APP + cMWS design.
This test succeeded, making it the first demonstration the focal plane wavefront sensing capabilities of the cMWS on-sky. 
A second run in 2017 used a simplified instrument, replacing the SLM for a vAPP.
Two instrument properties were taken into account for the design of the vAPP. 
First, the ALPAO deformable mirror with 97 actuators, which has a limited control radius.
Second, the AO-loop speed was limited at 800 Hz by the camera frame rate.
Thus, the vAPP designs were optimized for $10^{-4}$ contrast between 2.5-6 $\lambda/D$, resulting in a phase pattern 0-$\pi$ concentric rings.
A grating mask used to define the aperture with the vAPP.
With a narrow science band between 600 nm and 800 nm, the vAPP is a 1TR, resulting in a polarization leakage less than 6\%. 
The intensity of the leakage at the IWA is lower than the design contrast, which enabled the use of a 360 vAPP without gratings. 
The vAPP was multiplexed with 20 Disk Harmonic holograms, resulting in an improved wavefront correction \cite{haffert2018}.
\subsubsection{Subaru/SCExAO/CHARIS}
\begin{figure}
    \centering
    \includegraphics[width=\linewidth]{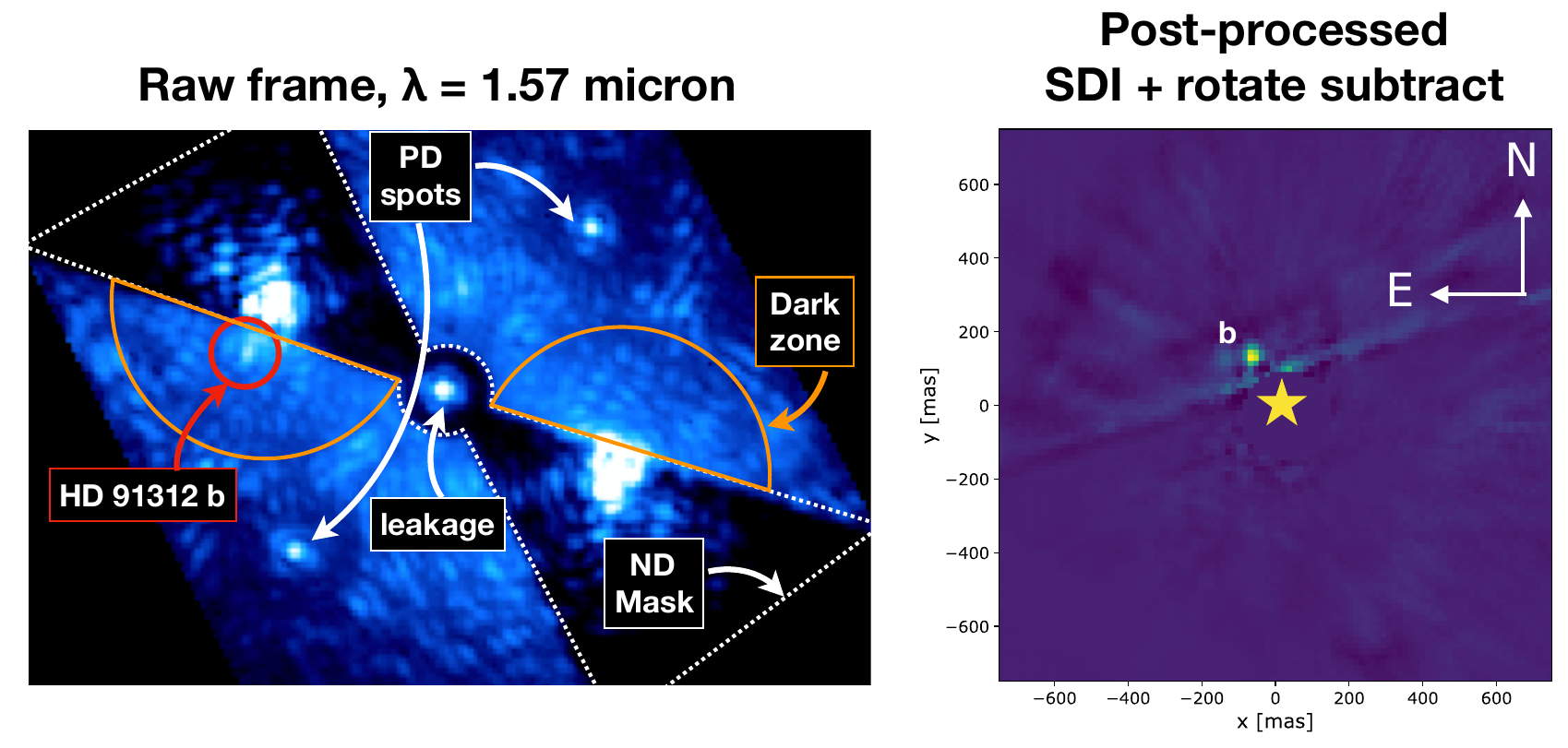}
    \caption{Observations of HD 91312 with the gvAPP with CHARIS. \textit{Left:} Image slice from the extracted data cube showing the raw PSF. {Right:} Final image of HD 91312, showing the companion clearly. The bright streak is a result of post-processing. The scale of both images is different.}
    \label{fig:HD91312}
\end{figure}

The Subaru Coronagraphic Extreme Adaptive Optics (SCExAO) \cite{Jovanovic2015} instrument is an ever-evolving high-contrast imaging system at the 8.2-m Subaru Telescope, with the aim of providing a platform for technology demonstration for next generation telescopes. A more recent, and already outdated, overview of the instrument can be found in \cite{lozi2020status}.
The gvAPP in SCExAO was installed in 2017 in the near-infrared arm (0.9-2.5 $\mu$m) \cite{doelman2017}.
The design of the gvAPP has a reduced grating frequency compared to previous designs to make the PSF fit in the SCExAO/CHARIS field of view (FOV) for the full bandwidth \cite{doelman2017}.
CHARIS is an integral field spectrograph with a resolution of $R\sim18.4$ for a bandwidth spanning the $J$, $H$, and $K$ bands (1.13-2.39 $\mu$m) and a FOV of $2"\times2"$ \cite{groff2017first}.
To fit these instrument specifications, the gvAPP was manufactured with a new liquid-crystal recipe with a diffraction efficiency greater than 97\% between 1-2.5 $\mu$m.
Two additional features of this gvAPP are the grating mask that is undersized with respect to the amplitude mask of SCExAO and defines the gvAPP pupil, and the first implementation of phase diversity spots.
At the inner working angle of 2.1 $\lambda/D$, the design contrast is $10^{-5}$, and the outer working angle is 11 $\lambda/D$, which is limited by the CHARIS FOV in K-band.\\
The gvAPP was installed together with a patterned ND-filter in the focal plane, which reduces the coronagraphic and leakage PSF intensity by a factor 10-50, depending on wavelength.
An image of the PSF with the patterned ND-filter is shown in Fig. \ref{fig:HD91312}.
Such a focal plane mask enables longer integration times and reduces the risk of saturation and resulting persistence of the sensitive NIR cameras.
The PD holograms are not blocked and can also be used as a photometric reference.
After installation, the first-light images of the gvAPP with CHARIS showed a PSF shift by $\sim 1$ arcsecond caused by the wedge of the optic.
A second wedged optic that is identical to the one used in the gvAPP was installed in a filter wheel downstream to successfully correct this shift. \\
Testing with the internal source showed promising results with a raw contrast $<10^{-3}$ at the inner working angle, and $<10^{-4}$ beyond 4.5 $\lambda/D$ \cite{bos2019}, demonstrating that the delivered gvAPP is within specifications and commensurate with the performance of the system and other coronagraphs.
However, on-sky testing resulted in raw-contrasts that were a factor 5-10 worse, caused by prominent diffraction structures present in the dark zones.
These structures are visible in Fig. \ref{fig:HD91312} as bright lines of speckles orthogonal to the flat side of the dark zone.
These structures are not present when using any other coronagraph present in SCExAO.
Inspection of pupil images with and without the gvAPP showed that these structures are not caused by pupil misalignment. 
In addition, the absence of these structures in the internal source data suggests that they are caused by optics upstream of SCExAO, e.g. the telescope mirrors or AO188, an upstream adaptive optics instrument.
Simulations show that annular phase offsets at the pupil edges can reproduce similar dark zone structures, which would not be noticed by other coronagraphs when a Lyot stop is used.
An annular phase structure can be seen in the on-sky wavefront reconstruction in \cite{bos2019} and it also has been observed with the pyramid wavefront sensor, but so far any attempt to correct these phase offsets did not result in a significant improvement of the dark zone contrast.\\
Despite the lower-than-expected raw-contrast, on-sky observations have been carried out with both the imager SCExAO/Chuckcam and the IFU SCExAO/CHARIS.
One of the first tests of the gvAPP was the observation of HD 91312 with CHARIS on 7 February 2018, for a total of integration time of 5 minutes 29 seconds.  
HD 91312 is a good candidate as it is a binary system, with a companion at a separation of 150 mas ($\sim$ 5 AU), which has a minimal mass of 80 M$_{Jup}$ \cite{borgniet2019}, and therefore is most likely a M-dwarf.
Because HD 91312 A is an A-star, the relative contrast to HD 91312 b in the near infrared is already favourable and becomes lower towards longer wavelengths.
The companion was detected in the raw frames, even with the diffraction structures present. 
Post-processing with spectral differential imaging (SDI) \cite{racine1999speckle} and rotate subtract \cite{otten2017} removed the diffraction structures.
The extracted PSF at $\lambda = 1.57\mu$m and the final post-processed image are shown in Fig. \ref{fig:HD91312}.
While the on-sky raw contrast is limited by the uncorrected diffraction structures, the SCExAO gvAPP is still valuable as a technology demonstration of focal-plane wavefront sensing.
The phase diversity between both coronagraphic PSFs, combined with the asymmetric pupil of SCExAO, resulted in the first demonstration of FPWFS with only the coronagraphic PSFs \cite{bos2019}.
The improvement of a factor two in raw contrast between 2-4 $\lambda/D$ is similar to other FPWFS techniques \cite{galicher2019}.
An additional improvement can be realised by implementing LDFC \cite{miller2019,miller2020spatial,bos2021submittedfirst}. 
However, it is unclear if the combination of both FPWFS techniques will resolve the issue with diffraction structures in the on-sky PSF. 
A future upgrade of AO188 could resolve the issue and enhance the gvAPP performance.
\subsubsection{FAST-CARMEN/HiCIBaS}
As is mentioned in the name, the High-Contrast Imaging Balloon System (HiCIBaS) is a balloon-borne telescope and is installed on the FAST-CARMEN gondola \cite{cote2018precursor}.
\begin{figure}
    \centering
    \includegraphics[width=0.88\linewidth]{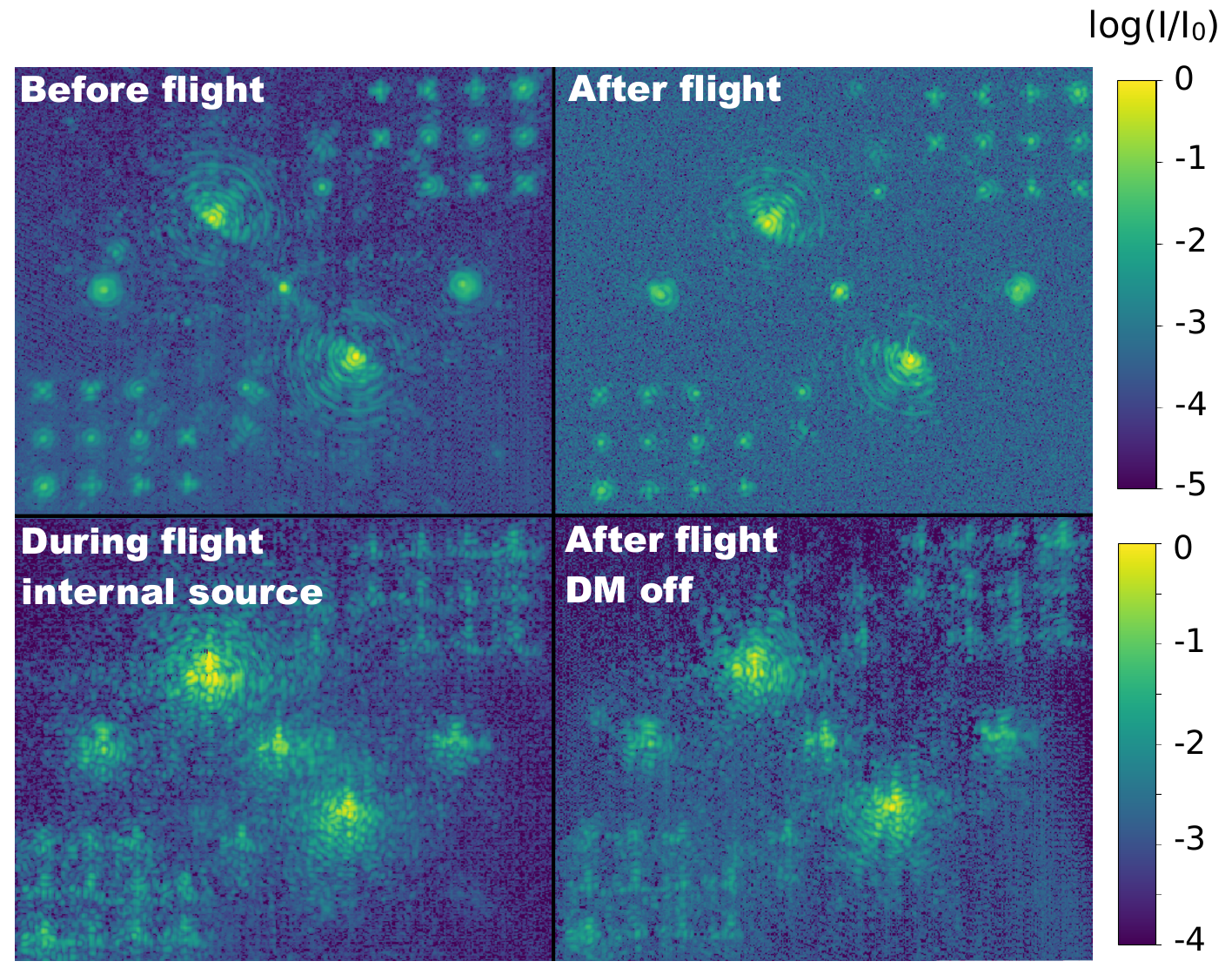}
    \caption{Measurements of the HiCIBaS gvAPP PSF (log$_{10}$ scale) before, during, and after the flight using the internal source of the HiCIBaS instrument. No optical component of HiCIBaS sustained major damage during the landing. The system was realigned after the flight.}
    \label{fig:hicibas_results}
\end{figure}
The main goals of this mission were to test new high-contrast imaging technology in a space-like environment and the training of five master students of Laval university \cite{thibault2019}. 
The scientific goal was to measure and characterize the wavefront distortions from high-altitude ($>$40km) turbulence, using low-order wavefront sensors (LOWFS) \cite{allain2018}.
The balloon launched August 25 2018 at 11:18 pm local time from the Timmins Stratospheric Balloon Base in Ontario, Canada.
The flight plan was to ascend to 40 km, and stay there for the duration of $\sim$ 12 hours.
A separate arm was created for high-contrast imaging. 
This arm would benefit from the pointing stabilization necessary for the LOWFS measurements.
The 37-element IRIS AO MEMS deformable mirror was installed to correct wavefront aberrations, however, it was not connected to the LOWFS to reduce complexity. 
The gvAPP was selected because the coronagraph is stable against residual pointing inaccuracies.
The design has a small dark zone from 2.1 to 8.5 $\lambda/D$ and a contrast of $10^{-6}$, delivering more extreme performance in the low-turbulance regime.
With a limited bandwidth between 0.83 and 0.88 $\mu$m, a 1TR gvAPP was sufficient.
Also, the gvAPP was designed to provide complementary wavefront sensor information using a 12-mode Zernike basis cMWS or PD holograms. 
The performance of the LOWFS and the gvAPP FPWFS were successfully tested close-loop in the system before the flight.
The PSF before the flight is shown in Fig. \ref{fig:hicibas_results}.
An internal laser source was used to show that both the cMWS and the PD holograms worked.\\
While the launch and flight happened according to the flight plan, an unfortunate mechanical failure resulted in the inability to operate the telescope. 
As a result, the telescope never pointed at a celestial source.
An internal source was present in the flight instrument case of such a failure, which was used to record the gvAPP PSFs.
It was also discovered that the deformable mirror did not respond to commands in flight and was stuck in a random, non-flattened state.
Extreme noise from electronic crosstalk resulted in low-quality images of the PSF.
The resulting images are not useful to characterize the gvAPP performance during flight, see  Fig. \ref{fig:hicibas_results}.
The balloon and payload were recovered after the flight. 
The high-contrast imaging system, including the gvAPP, were still operational. 
Post-launch images of the gvAPP PSF do not show any significant deviation from pre-launch performance, and the flight PSF is similar to the after-flight PSF with the DM off. 
We note that both comparisons are qualitative, and have not been fully characterized. 
We conclude that the gvAPP or similar LC coronagraph can operate in a space-like environment. 
\subsubsection{Magellan/MagAO-X}
The MagAO-X instrument is a completely different instrument than MagAO, operating in the visible and near-infrared with an extreme AO (ExAO) system \cite{males2018}.
The key science goal of MagAO-X is high-contrast imaging of accreting protoplanets at H$\alpha$ (656.28 nm).
To achieve this goal, MagAO-X will deliver high Strehls ($>$70\% at  H$\alpha$), and high contrasts ($<10^{-4}$) from $\sim$ 1 to 10 $\lambda/D$.
In addition, MagAO-X implements many of the lessons-learned from SCExAO, including the real-time controller CACAO\footnote{\url{https://github.com/cacao-org}}. 
The first phase (Phase I) aims to fully characterize and calibrate the instrument while obtaining the first science results. 
With non-optimal wavefront correction in phase I, the gvAPP was chosen as a robust coronagraph.
To accommodate wavefront sensing and good coronagraphic performance, the Strehl ratio of the design is lower than most other coronagraphs (40.7\%).
The coronagraph design creates a dark zone from 2.1-15 $\lambda/D$, with $10^{-5}$ contrast at the IWA, and the dark zones are elongated in the grating direction. 
Therefore, wavelength smearing up to 20\% does not compromise the dark zone towards the edges.\\
The gvAPP pupil was specifically altered with a thicker spider to be asymmetric with the goal of enabling focal-plane WFS using the coronagraphic PSFs.
As a result, the gvAPP design employs three ways of wavefront sensing, i.e. PD, cMWS, and AP-WFS, in addition to LDFC for wavefront control. 
Each of these wavefront sensing modes have been tested separately using a test plate with different gvAPP designs \cite{miller2018development,miller2019}.
The wavefront sensing capabilities enable better instrument calibration by accurately measuring non-common path aberrations (NCPA), which are more severe in the visible.
MagAO-X employs a separate DM that is dedicated to correct NCPA, which will be fed directly by the focal-plane wavefront sensors of the gvAPP.
FPWFS can be done at much greater speeds with MagAO-X because a laser-cut mask in the intermediate focal plane separates the dark zones from the bright field.
This mask transmits the dark holes to the EMCCD science camera and reflects the stellar bright field back to a second EMCCD that will act as the dedicated FPWFS.
The cMWS was especially useful for alignment of the off-axis paraboloids, providing direct visual feedback on the first 9 low-order Zernike modes.
For on-sky operations, a likely strategy would be to use PD, cMWS and AP-WFS to minimize the non-common path aberrations before observing and use LDFC to remove any drifts.\\
MagAO-X had first light in 2019, where it demonstrated a Strehl ratio of 46\%, and a 5$\sigma$ post-processed gvAPP contrast of $10^{-4}$ at 3 $\lambda/D$ at 900 nm (z') \cite{males2020magao}.
These results are promising for discovering exoplanets during formation through direct detection of $H\alpha$ from accretion of hydrogen onto these protoplanets.
Applying the Massive Accreting Gap (MAG) protoplanet model\cite{close2020separation} to the observed first light MagAO-X contrasts results in a predicted maximum yield of 46$\pm$7 planets from 19 stars\cite{close2020prediction}.
\subsubsection{LBT/LMIRcam/ALES}
The Large Binocular Telescope Mid-Infrared camera (LMIRcam) is an imager/spectrograph that operates behind the Large Binocular Telescope Interferometer (LBTI) \cite{skrutskie2010}. 
The LBT is a unique telescope with two 8.4-m primary mirrors mounted next to each other.
Both apertures can be combined with LBTI and create fringes on LMIRcam, resulting in a 22.8-m baseline.
The LBT and LMIRcam are ideal for $L$ and $M$ band observations with a minimal number of warm optics and a deformable secondary mirror.
This makes LBTI/LMIRCAM the most sensitive infrared instrument in the Northern hemisphere. 
\begin{figure}
    \centering
    \includegraphics[width=\linewidth]{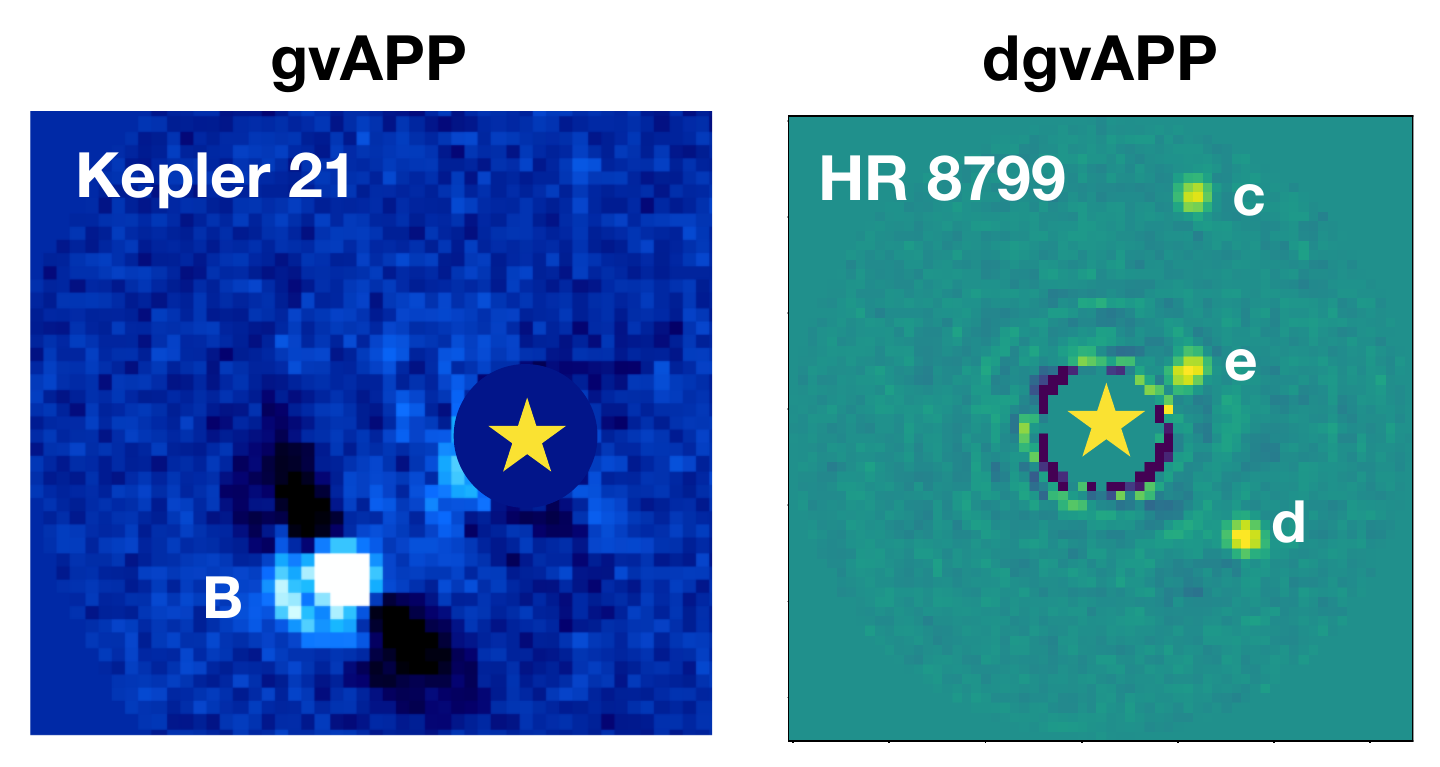}
    \caption{Detection of Kepler 21B using the gvAPP at LBT/LMIRcam ($\Delta m = 3.9$ and the inner three HR 8799 planets ($\Delta m = \sim 10$) using the dgvAPP at LMIRcam/ALES. }
    \label{fig:LBT_results}
\end{figure}
We take advantage of this opportunity with two different vAPPs, a grating vAPP (1.8-15 $\lambda/D$) and a double-grating vAPP (2.7-15 $\lambda/D$).
The designs are pushing the limits in inner working angle, and the gvAPP also incorporates focal-plane wavefront sensing with PD holograms and photometric/astrometric reference spots.
Their more extreme designs are enabled by the small secondary and the extremely thin secondary support structure.
The optic contains two vAPPs, one for each aperture.
Their beams can be imaged separately, partially combined with overlapping fringing holograms, or completely overlapped to produce fringes for nulling. \\
LMIRCam was augmented in 2015 with the Arizona Lenslets for Exoplanet Spectroscopy (ALES) \cite{skemer2015first}, an integral field spectrometer (IFS).
After a recent upgrade, ALES facilitates 2 to 5 $\mu$m low-resolution (R$\sim$35) spectroscopy of directly imaged gas-giant exoplanets \cite{skemer2018ales}.
Currently, it is the only IFS operating in the $L$ and $M$ bands, containing broad molecular features (CO, CH$_4$), and also covers the 3.1 $\mu$m ice feature, 3.3 $\mu$m PAH feature, and 4.1 $\mu$m Br-alpha emission line.
Therefore, low-resolution spectroscopy can provide a unique window for gas giants in the L-T transition, where their atmospheres transition from CO-rich and cloudy (L-type), to CH$_4$-rich and relatively cloud free (T-type).\\
Both vAPPs installed in LMIRcam are available for ALES, and, together with the annular groove phase mask (AGPM) coronagraph, they are the only coronagraphs installed in LMIRcam/ALES that cover the full wavelength range.
Both vAPPs do have significant absorption (up to 100\%) between 3.15 and 3.45 $\mu$m from glue and the liquid-crystals.
The outer working angle of both coronagraphs is larger than the FOV of ALES for wavelengths larger than $\sim 3$ $\mu$m.
For the gvAPP only one of the two coronagraphic PSFs can be fitted inside the FOV, and the wavelength smearing moves it across the detector.
While this is not ideal, the gvAPP provides a much smaller inner working angle compared to the dgvAPP.
The dgvAPP provides a stable central PSF and is more sensitive than the gvAPP due to the recombination of both coronagraphic PSFs. 
Moreover, as it provides a 360 degree dark zone it is much more user friendly and intuitive. \\
Tests with the internal source show that both vAPPs are operating as expected. 
Only a few observations have been carried out, but already show the great promise of these vAPPs, see Fig. \ref{fig:LBT_results}.
The results on the HR 8799 system will be discussed in more detail in a future paper.
The unique feature of the vAPPs is the presence of the stellar PSF on the detector.
This is particularly interesting for photometric monitoring of exoplanets.
For example, monitoring the stellar PSF can be used to reduce the impact of variability in AO performance during the observing run. 
In addition, an upgraded data reduction technique has been developed to remove the thermal background with great accuracy using the science frames themselves.
Nodding the telescope to characterize the thermal background with off-source measurements is still necessary, however at much lower cadence, increasing the on-source time.
To summarize, there are still many opportunities with the vAPPs in LMIRcam/ALES.
\subsubsection{VLT/ERIS}
The Enhanced Resolution Imager and Spectrograph (ERIS) will be installed in the Cassegrain focus of UT4 at the Very Large Telescope in the near future \cite{amico2012design,davies2018eris}.
ERIS is a 1-5 $\mu$m instrument with an imager (NIX) operating from $J$-$M_p$ band and an IFS (SPIFFIER) in $J$-$K$ band with R$\sim$8000. 
ERIS/NIX will have a vortex coronagraph and a gvAPP, while ERIS/SPIFFIER is non-coronagraphic. 
The ERIS gvAPP has a dark zone from 2.2-15 $\lambda/D$, which has been widened to accommodate wavelength smearing, and the APP phase is multiplexed with two photometric/astrometric reference spots \cite{kenworthy2018high}. 
The pupil has an asymmetric central obscuration because the (warm) tertiary mirror (M3) will be moved aside to illuminate the Cassegrain focus where ERIS is residing.
While the asymmetry is small, it is still possible to estimate NCPA with FPWFS \cite{bos2019} if the software is implemented.
The gvAPP has been manufactured in 2018, and was later tested in the near-infrared test bench of the group for Exoplanet \& Habitability at ETH Zurich \cite{boehle2018cryogenic}.
The setup imaged the gvAPP PSF at 3.8 $\mu$m and was limited by background noise to $\sim 10^{-3}$.
Since then an updated setup has been realized and the gvAPP was tested to the $\sim 10^{-5}$ level \cite{boehle2021submitted}. 
These measurements demonstrate the ERIS/gvAPP operates as expected, within the limits of the test bench. \\
While ERIS/NIX is an imager, it will have a long-slit spectroscopic mode with R$\sim$850, covering the full $L$ band.
With the gvAPP absorption feature between 3.15-3.45 $\mu$m, direct detection of methane will be more difficult.
For atmospheres with elevated methane levels, e.g. T-type brown dwarfs, the spectral slope beyond 3.5 $\mu$m can be used to measure methane levels.
Another unique opportunity of ERIS/NIX is enabled by the combination of a sensitive wavefront sensor and the powerful (20W) laser guide star.
The laser guide star allows observations of much fainter science targets, which will greatly increase the number of potential targets accessible to this new instrument.
\subsubsection{ELT/METIS and ELT/MICADO}
The Extremely Large Telescope (ELT) \cite{gilmozzi2007european} will revolutionize direct imaging of exoplanets because of its primary mirror diameter of 39 meters.
The supreme resolution will be fully exploited by first generation instruments.
With five times the resolution of the VLT and 25 times the sensitivity, the ELT will provide new insights in planet formation, protoplanetary disks, planet-disk interaction, planetary atmospheres and planet evolution.
Both the Mid-infrared ELT Imager and Spectrograph (METIS) \cite{brandl2016status} and the Multi-Adaptive Optics Imaging CamerA for Deep Observations (MICADO) \cite{davies2016micado} have exoplanet detection and characterization as one of the main science goals.
To this end, both instruments have a gvAPP as a baseline coronagraph. \\
METIS covers the science bands between 3-14 $\mu$m, and has a high-resolution IFS mode with $R\sim100,000$ in $L$ and $M$ band, in addition to an imaging mode from $L$ to $N$ band. 
This high spectral resolution is achieved over a large FOV of 0.5 by 1.0 arcseconds, which is made possible by an image slicer that divides the field of view into multiple strips on a slicer mirror. 
Simulations show that METIS will detect and characterize multiple planets found with radial-velocity studies \cite{quanz2015direct}.
Two gvAPPs with different dark zone designs will be installed in the IFS and the imager in $L$ and $M$ band \cite{kenworthy2018review}. 
A specific downside of the gvAPP is that the star light is not blocked, leading to high intensities in the IFS focal plane. 
This can be prevented by changing the position of the star in the field of view using the chopper.
Due to this offset only the dark zone of the gvAPP PSF is imaged onto the image slicer. 
The ELT pupil is not favourable for coronagraph design, with the large central obscuration.
The METIS pupil is a cold stop and has additional thick spiders, complicating the design process further.
The METIS gvAPP design therefore has a larger inner working angle of 2.5 $\lambda$/D to keep the Strehl above 65\%.
Moreover, the gvAPP has a more moderate contrast towards the inner working angle, going down from a few times $10^{-4}$ at 3 $\lambda/D$ to $10^{-6}$ beyond 6 $\lambda/D$.
Simulations show that this does not affect the performance of the coronagraph compared to other baselined coronagraphs like the AGPM, as NCPA, residual wavefront aberrations, and missing mirror segments of the primary will limit the performance at the smallest separations.
In addition, the AGPM is quite sensitive to tip-tilt which can be significant for nodding and high frequency structural vibrations \cite{gluck2016simulation}. \\
MICADO, the European ELT first-light imager \cite{davies2018}, covers the science bands between 0.8-2.4 $\mu$m and will benefit from both the M4 single-conjugate AO system \cite{clenet2018} and the dedicated multi-conjugate AO system instrument (MAORY) \cite{diolaiti2016maory}. 
With diffraction limited performance, MICADO will deliver extremely high-resolution imagery to study the inner regions of planetary systems \cite{baudoz2019}. 
The high contrast imaging modes will comprise two classical Lyot coronagraphs \cite{perrot2018}, one gvAPP and two non redundant masks \cite{Lacour2014}. 
One challenge in designing the high contrast imaging modes of MICADO comes from the fact that they have to be limited to one focal plane upstream the atmospheric dispersion corrector, and one downstream pupil plane. 
While the atmospheric dispersion might have an impact on the performance of the smallest Lyot coronagraph in broadband, gvAPP is not sensitive to this effect and will thus offer high performance even at high airmass.
The MICADO gvAPP design is optimized between 2.5 $\lambda/D$ and 20 $\lambda/D$ with a square shape. 
On-sky expected contrast performance reaches $2 \times 10^{-5}$ at 3.5 $\lambda/D$ in K band with a Strehl ratio of 70\%. For bright nearby stars, the starlight could saturate the detector for the shortest exposure time, and the use of a neutral density filter might be necessary.

%
\section{Current status and future developments}
\label{sec:observing}
With the vAPP successfully installed in many different telescopes, the focus for the vAPP developments is now transitioning from testing and commissioning the technology, to science observations.
Much of the time spent on vAPPs so far has been oriented towards technology development and design optimization. 
However, the on-sky technology demonstration, observation planning, and data reduction are critical to achieve competitive science results.\\
The majority of the current on-sky data with a vAPP have been taken with the MagAO, LEXI and SCExAO instruments. 
Bright targets were selected to provide optimal AO correction for technology demonstration of the vAPP and FPWFS. 
While the technology demonstration was often successful and benefited from test observations and consequent understanding and iterations, time awarded to science observations has been limited, and conditions for these observations have been sub-optimal.
Attempts on more challenging targets, e.g. HD 206893 with SCExAO, resulted in non-detections.
\textcolor{black}{Due to the sub-optimal circumstances, limited available data and sub-optimal data reduction methods, it is difficult to gauge the performance with respect to other coronagraphs.}
We tried bridging this gap by pursuing targets with intermediate contrast by looking into stellar multiplicity and to understand the detection limits.
This science case was awarded a few nights at the LBT and Subaru, most of which were lost due to bad weather. 
Currently the vAPPs of the LBT have yielded the best results, recovering HR 8799 c,d,e. \\
To fully exploit the opportunities of the ERIS, MICADO and METIS gvAPPs, getting more experience with on-sky observations is paramount.
We note that these instruments have had a gvAPP baselined in the instrument design, and have dedicated engineering time planned to resolve any issues.
Moreover, of all the vAPPs that are currently installed, only the SCExAO gvAPP did not reach the expected performance on-sky.
The diffraction structures in the SCExAO dark zones are not caused by the gvAPP itself because the lab results do not show these structures. 
So, it is likely caused by uncorrected aberrations from the optics upstream of the internal source injection, which is unlikely for the ERIS instrument that is installed in the Cassegrain focus.
For METIS and MICADO the vAPPs will be tested extensively and problems should be identified during this process.
We conclude that future efforts should focus on observation planning and a standardized vAPP data reduction pipeline.\\
For both observation planning and a data reduction pipeline a preliminary version already exists, as discussed in Section \ref{sec:observe_data}. 
The vAPP observation preparation tool does include most vAPPs that are currently available, however, it does not provide a proper indication whether a companion will be in the dark zone.
In future upgrades we will add this functionality, in addition to showing the field rotation of the companion in the dark zones of on-sky PSFs.
Ultimately, this could be used to calculate which days of a semester are optimal for observations of known objects to help proposal writing.\\
The data reduction pipelines are currently all custom for each vAPP.
A unification of these pipelines would centralize all efforts on vAPP data reduction and provide a standardized pipeline that can be used by anyone with vAPP data.
In addition, it would increase efficiency of implementing new algorithms like rotate-subtract or TRAP.
To minimize duplication of previous efforts, it is essential that the pipeline can also port to existing post-processing pipelines like \textcolor{black}{the Vortex Image Processing package \cite{gonzalez2017vip} or PynPoint} \cite{stolker2019pynpoint}. 
Lastly, these pipelines should include methods to use capitalize on the astrometric and photometric holographic PSFs. \\
Future technical developments should be towards enabling broadband imaging with the vAPP. 
The main challenge to overcome is a (super)achromatic QWP that adds minimal wavefront aberration and ghosts to the system \cite{bos2018fully}. 
By adding another QWP before the vAPP, it is also possible to combine the vAPP with imaging polarimetry \cite{snik2014combining, bos2018fully}, another powerful tool to distinguish astronomical signals from speckle noise \cite{kuhn2001imaging}. 
Adding multiple coronagraphic PSFs per polarization state can then simplify more advanced polarimetric vAPP designs \cite{bos2020new}. 
Another potential interesting avenue to explore is the combination of the vAPP with multi-color holography \cite{doelman2019multi}, as that would completely eliminate leakage problems \cite{bos2020new}. 
However, this implementation is more difficult to implement in an instrument and puts strong requirements on the shape of the pupil.  
\section{Conclusion}
The vector-apodizing phase plate coronagraph is a versatile single-optic pupil-plane coronagraph that is easily adapted to any telescope aperture.
It delivers good contrast (<10$^{-4}$, limited by AO performance) at small inner working angles ($\sim 2 \lambda/D$), it can host many different focal-plane wavefront sensing techniques, and can include astrometric and photometric reference PSFs.
The manufacturing using direct-write for liquid-crystal technology enables extreme and accurate patterning and the mult-twist retarder technology provides excellent efficiencies ($>96\%$) for bandwidths up to 100\% between 0.55 $\mu$m and 5 $\mu$m.
Because of these properties, the vAPP has now been installed in 6 different instruments.
Measurements with the internal source of these instruments demonstrate that the vAPPs are operating as expected.
The grating of gvAPPs and their D-shaped dark zone make observing and data reduction more complicated compared to focal-plane coronagraphs.
Future efforts should focus on software for observation planning and data reduction to fully exploit the opportunities provided by future instruments that have the vAPP baselined, i.e. VLT/ERIS, ELT/METIS, and ELT/MICADO.

\section{Acknowledgements}
The research of David Doelman, Steven Bos, and Frans Snik leading to these results has received funding from the European Research Council under ERC Starting Grant agreement 678194 (FALCONER). Ben Sutlieff is fully supported by the Netherlands Research School for Astronomy (NOVA). 
MagAO-X was funded by NSF MRI Award \#1625441. 
KMM's work is supported by the NASA Exoplanets Research Program (XRP) by cooperative agreement NNX16AD44G.
This research made use of HCIPy, an open-source object-oriented framework written in Python for performing end-to-end simulations of high-contrast imaging instruments \cite{por2018}. 
This paper is based on work funded by NSF Grant 1608834.

\section{Disclosures}
The authors declare no conflicts of interest.
\bibstyle{osajnl}
\bibliography{vAPP_overview}

\begin{thebibliography}{100}
\newcommand{\enquote}[1]{``#1''}

\bibitem{lyot1939study}
B.~Lyot, \enquote{The study of the solar corona and prominences without
  eclipses (george darwin lecture, 1939),} {\protect\JournalTitle{Monthly
  Notices of the Royal Astronomical Society}} \textbf{99}, 580 (1939).

\bibitem{mawet2012review}
D.~Mawet, L.~Pueyo, P.~Lawson, L.~Mugnier, W.~Traub, A.~Boccaletti, J.~T.
  Trauger, S.~Gladysz, E.~Serabyn, J.~Milli \emph{et~al.}, \enquote{Review of
  small-angle coronagraphic techniques in the wake of ground-based
  second-generation adaptive optics systems,} in \emph{Space Telescopes and
  Instrumentation 2012: Optical, Infrared, and Millimeter Wave,}  vol. 8442
  (International Society for Optics and Photonics, 2012), p. 844204.

\bibitem{ruane2018}
G.~Ruane, A.~Riggs, J.~Mazoyer, E.~Por, M.~N'Diaye, E.~Huby, P.~Baudoz,
  R.~Galicher, E.~Douglas, J.~Knight \emph{et~al.}, \enquote{Review of
  high-contrast imaging systems for current and future ground-and space-based
  telescopes i: coronagraph design methods and optical performance metrics,} in
  \emph{Space Telescopes and Instrumentation 2018: Optical, Infrared, and
  Millimeter Wave,}  vol. 10698 (International Society for Optics and
  Photonics, 2018), p. 106982S.

\bibitem{nielsen2019gemini}
E.~L. Nielsen, R.~J. De~Rosa, B.~Macintosh, J.~J. Wang, J.-B. Ruffio,
  E.~Chiang, M.~S. Marley, D.~Saumon, D.~Savransky, S.~M. Ammons \emph{et~al.},
  \enquote{The gemini planet imager exoplanet survey: giant planet and brown
  dwarf demographics from 10 to 100 au,} {\protect\JournalTitle{The
  Astronomical Journal}} \textbf{158}, 13 (2019).

\bibitem{vigan2020sphere}
A.~Vigan, C.~Fontanive, M.~Meyer, B.~Biller, M.~Bonavita, M.~Feldt,
  S.~Desidera, G.-D. Marleau, A.~Emsenhuber, R.~Galicher \emph{et~al.},
  \enquote{The sphere infrared survey for exoplanets (shine). iii. the
  demographics of young giant exoplanets below 300 au with sphere,}
  {\protect\JournalTitle{arXiv preprint arXiv:2007.06573}}  (2020).

\bibitem{avenhaus2018disks}
H.~Avenhaus, S.~P. Quanz, A.~Garufi, S.~Perez, S.~Casassus, C.~Pinte, G.~H.-M.
  Bertrang, C.~Caceres, M.~Benisty, and C.~Dominik, \enquote{Disks around t
  tauri stars with sphere (dartts-s). i. sphere/irdis polarimetric imaging of
  eight prominent t tauri disks,} {\protect\JournalTitle{The Astrophysical
  Journal}} \textbf{863}, 44 (2018).

\bibitem{codona2006}
J.~Codona, M.~Kenworthy, P.~M. Hinz, J.~R.~P. Angel, and N.~Woolf, \enquote{A
  high-contrast coronagraph for the mmt using phase apodization: design and
  observations at 5 microns and 2 $\lambda$/d radius,} in \emph{Ground-based
  and Airborne Instrumentation for Astronomy,}  vol. 6269 (International
  Society for Optics and Photonics, 2006), p. 62691N.

\bibitem{kenworthy2007}
M.~A. Kenworthy, J.~L. Codona, P.~M. Hinz, J.~R.~P. Angel, A.~Heinze, and
  S.~Sivanandam, \enquote{First on-sky high-contrast imaging with an apodizing
  phase plate,} {\protect\JournalTitle{The Astrophysical Journal}}
  \textbf{660}, 762 (2007).

\bibitem{quanz2010}
S.~P. Quanz, M.~R. Meyer, M.~A. Kenworthy, J.~H. Girard, M.~Kasper, A.-M.
  Lagrange, D.~Apai, A.~Boccaletti, M.~Bonnefoy, G.~Chauvin \emph{et~al.},
  \enquote{First results from very large telescope naco apodizing phase plate:
  4 $\mu$m images of the exoplanet $\beta$ pictoris b,}
  {\protect\JournalTitle{The Astrophysical Journal Letters}} \textbf{722}, L49
  (2010).

\bibitem{Quanz13}
S.~P. {Quanz}, A.~{Amara}, M.~R. {Meyer}, M.~A. {Kenworthy}, M.~{Kasper}, and
  J.~H. {Girard}, \enquote{A young protoplanet candidate embedded in the
  circumstellar disk of hd 100546,} {\protect\JournalTitle{The Astrophysical
  Journal Letters}} \textbf{766}, L1 (2013).

\bibitem{meshkat2015}
T.~Meshkat, V.~P. Bailey, K.~Y. Su, M.~A. Kenworthy, E.~E. Mamajek, P.~M. Hinz,
  and P.~S. Smith, \enquote{Searching for planets in holey debris disks with
  the apodizing phase plate,} {\protect\JournalTitle{The Astrophysical
  Journal}} \textbf{800}, 5 (2015).

\bibitem{pancharatnam1956}
S.~Pancharatnam, \enquote{Generalized theory of interference and its
  applications,} {\protect\JournalTitle{Proceedings of the Indian Academy of
  Sciences - Section A}} \textbf{44}, 398--417 (1956).

\bibitem{berry1987}
M.~V. Berry, \enquote{The adiabatic phase and pancharatnam's phase for
  polarized light,} {\protect\JournalTitle{Journal of Modern Optics}}
  \textbf{34}, 1401--1407 (1987).

\bibitem{snik2012}
F.~Snik, G.~Otten, M.~Kenworthy, M.~Miskiewicz, M.~Escuti, C.~Packham, and
  J.~Codona, \enquote{The vector-app: a broadband apodizing phase plate that
  yields complementary psfs,} in \emph{Modern Technologies in Space-and
  Ground-based Telescopes and Instrumentation II,}  vol. 8450 (International
  Society for Optics and Photonics, 2012), p. 84500M.

\bibitem{otten2014}
G.~P. Otten, F.~Snik, M.~A. Kenworthy, M.~N. Miskiewicz, M.~J. Escuti, and
  J.~L. Codona, \enquote{The vector apodizing phase plate coronagraph:
  prototyping, characterization and outlook,} in \emph{Advances in Optical and
  Mechanical Technologies for Telescopes and Instrumentation,}  vol. 9151
  (International Society for Optics and Photonics, 2014), p. 91511R.

\bibitem{miskiewicz2014}
M.~N. Miskiewicz and M.~J. Escuti, \enquote{Direct-writing of complex liquid
  crystal patterns,} {\protect\JournalTitle{Optics Express}} \textbf{22},
  12691--12706 (2014).

\bibitem{komanduri2012}
R.~K. Komanduri, J.~Kim, K.~F. Lawler, and M.~J. Escuti, \enquote{Multi-twist
  retarders for broadband polarization transformation,} in \emph{Emerging
  Liquid Crystal Technologies VII,}  vol. 8279 (International Society for
  Optics and Photonics, 2012), p. 82790E.

\bibitem{komanduri2013}
R.~K. Komanduri, K.~F. Lawler, and M.~J. Escuti, \enquote{Multi-twist
  retarders: broadband retardation control using self-aligning reactive liquid
  crystal layers,} {\protect\JournalTitle{Optics Express}} \textbf{21},
  404--420 (2013).

\bibitem{bos2018fully}
S.~P. Bos, D.~S. Doelman, J.~de~Boer, E.~H. Por, B.~Norris, M.~J. Escuti, and
  F.~Snik, \enquote{Fully broadband vapp coronagraphs enabling polarimetric
  high contrast imaging,} in \emph{Advances in Optical and Mechanical
  Technologies for Telescopes and Instrumentation III,}  vol. 10706
  (International Society for Optics and Photonics, 2018), p. 107065M.

\bibitem{otten2014a}
G.~P. Otten, F.~Snik, M.~A. Kenworthy, M.~N. Miskiewicz, M.~J. Escuti, and
  J.~L. Codona, \enquote{The vector apodizing phase plate coronagraph:
  prototyping, characterization and outlook,} in \emph{Advances in Optical and
  Mechanical Technologies for Telescopes and Instrumentation,}  vol. 9151
  (International Society for Optics and Photonics, 2014), p. 91511R.

\bibitem{otten2017}
G.~P. Otten, F.~Snik, M.~A. Kenworthy, C.~U. Keller, J.~R. Males, K.~M.
  Morzinski, L.~M. Close, J.~L. Codona, P.~M. Hinz, K.~J. Hornburg
  \emph{et~al.}, \enquote{On-sky performance analysis of the vector apodizing
  phase plate coronagraph on magao/clio2,} {\protect\JournalTitle{The
  Astrophysical Journal}} \textbf{834}, 175 (2017).

\bibitem{bos2019}
S.~P. Bos, D.~S. Doelman, J.~Lozi, O.~Guyon, C.~U. Keller, K.~L. Miller,
  N.~Jovanovic, F.~Martinache, and F.~Snik, \enquote{Focal-plane wavefront
  sensing with the vector-apodizing phase plate,}
  {\protect\JournalTitle{Astronomy \& Astrophysics}} \textbf{632}, A48 (2019).

\bibitem{haffert2018}
S.~Haffert, M.~Wilby, C.~Keller, I.~Snellen, D.~Doelman, E.~Por, M.~van Kooten,
  S.~Bos, and J.~Wardenier, \enquote{On-sky results of the leiden exoplanet
  instrument (lexi),} in \emph{Adaptive Optics Systems VI,}  vol. 10703
  (International Society for Optics and Photonics, 2018), p. 1070323.

\bibitem{doelman2020}
D.~S. Doelman, E.~H. Por, G.~Ruane, M.~J. Escuti, and F.~Snik,
  \enquote{Minimizing the polarization leakage of geometric-phase coronagraphs
  with multiple grating pattern combinations,}
  {\protect\JournalTitle{Publications of the Astronomical Society of the
  Pacific}} \textbf{132}, 045002 (2020).

\bibitem{codona2004}
J.~L. Codona and R.~Angel, \enquote{Imaging extrasolar planets by stellar halo
  suppression in separately corrected color bands,} {\protect\JournalTitle{The
  Astrophysical Journal Letters}} \textbf{604}, L117 (2004).

\bibitem{kostinski2005}
A.~B. Kostinski and W.~Yang, \enquote{Pupil phase apodization for imaging of
  faint companions in prescribed regions,} {\protect\JournalTitle{Journal of
  Modern Optics}} \textbf{52}, 2467--2474 (2005).

\bibitem{keller2016novel}
C.~U. Keller, \enquote{Novel instrument concepts for characterizing directly
  imaged exoplanets,} in \emph{Ground-based and Airborne Instrumentation for
  Astronomy VI,}  vol. 9908 (International Society for Optics and Photonics,
  2016), p. 99089V.

\bibitem{carlotti2013}
A.~Carlotti, \enquote{Apodized phase mask coronagraphs for arbitrary
  apertures,} {\protect\JournalTitle{Astronomy \& Astrophysics}} \textbf{551},
  A10 (2013).

\bibitem{por2017}
E.~H. Por, \enquote{Optimal design of apodizing phase plate coronagraphs,} in
  \emph{Techniques and Instrumentation for Detection of Exoplanets VIII,}  vol.
  10400 (International Society for Optics and Photonics, 2017), p. 104000V.

\bibitem{cantalloube2020}
{Cantalloube, F.}, {Farley, O. J. D.}, {Milli, J.}, {Bharmal, N.}, {Brandner,
  W.}, {Correia, C.}, {Dohlen, K.}, {Henning, Th.}, {Osborn, J.}, {Por, E.},
  {Su\'arez Valles, M.}, and {Vigan, A.}, \enquote{Wind-driven halo in
  high-contrast images - i. analysis of the focal-plane images of sphere,}
  {\protect\JournalTitle{A\&A}} \textbf{638}, A98 (2020).

\bibitem{bloemhof2001}
E.~Bloemhof, R.~Dekany, M.~Troy, and B.~Oppenheimer, \enquote{Behavior of
  remnant speckles in an adaptively corrected imaging system,}
  {\protect\JournalTitle{The Astrophysical Journal Letters}} \textbf{558}, L71
  (2001).

\bibitem{wagner2020}
K.~Wagner, J.~Stone, R.~Dong, S.~Ertel, D.~Apai, D.~Doelman, A.~Bohn,
  J.~Najita, S.~Brittain, M.~Kenworthy \emph{et~al.}, \enquote{First images of
  the protoplanetary disk around pds 201,} {\protect\JournalTitle{The
  Astronomical Journal}} \textbf{159}, 252 (2020).

\bibitem{gurobi}
L.~Gurobi~Optimization, \enquote{Gurobi optimizer reference manual,}  (2020).

\bibitem{soummer2007}
R.~Soummer, L.~Pueyo, A.~Sivaramakrishnan, and R.~J. Vanderbei, \enquote{Fast
  computation of lyot-style coronagraph propagation,}
  {\protect\JournalTitle{Optics Express}} \textbf{15}, 15935--15951 (2007).

\bibitem{gerchberg1972practical}
R.~W. Gerchberg, \enquote{A practical algorithm for the determination of phase
  from image and diffraction plane pictures,} {\protect\JournalTitle{Optik}}
  \textbf{35}, 237--246 (1972).

\bibitem{por2018}
E.~H. Por, S.~Y. Haffert, V.~M. Radhakrishnan, D.~S. Doelman, M.~van Kooten,
  and S.~P. Bos, \enquote{High contrast imaging for python (hcipy): an
  open-source adaptive optics and coronagraph simulator,} in \emph{Adaptive
  Optics Systems VI,}  vol. 10703 (International Society for Optics and
  Photonics, 2018), p. 1070342.

\bibitem{dong2012}
S.~Dong, T.~Haist, and W.~Osten, \enquote{Hybrid wavefront sensor for the fast
  detection of wavefront disturbances,} {\protect\JournalTitle{Applied optics}}
  \textbf{51}, 6268--6274 (2012).

\bibitem{wilby2017}
M.~J. Wilby, C.~U. Keller, F.~Snik, V.~Korkiakoski, and A.~G. Pietrow,
  \enquote{The coronagraphic modal wavefront sensor: a hybrid focal-plane
  sensor for the high-contrast imaging of circumstellar environments,}
  {\protect\JournalTitle{Astronomy \& Astrophysics}} \textbf{597}, A112 (2017).

\bibitem{doelman2018}
D.~S. Doelman, P.~Tuthill, B.~Norris, M.~J. Wilby, E.~Por, C.~U. Keller, M.~J.
  Escuti, and F.~Snik, \enquote{Multiplexed holographic aperture masking with
  liquid-crystal geometric phase masks,} in \emph{Optical and Infrared
  Interferometry and Imaging VI,}  vol. 10701 (International Society for Optics
  and Photonics, 2018), p. 107010T.

\bibitem{bos2020new}
S.~P. Bos, D.~S. Doelman, K.~L. Miller, and F.~Snik, \enquote{New concepts in
  vector-apodizing phase plate coronagraphy,} in \emph{Adaptive Optics Systems
  VII,}  vol. 11448 (International Society for Optics and Photonics, 2020), p.
  114483W.

\bibitem{bos2020}
S.~P. Bos, \enquote{Vector speckle grid: instantaneous incoherent speckle grid
  for high-precision astrometry and photometry in high-contrast imaging,}
  {\protect\JournalTitle{arXiv preprint arXiv:2005.08751}}  (2020).

\bibitem{miller2019}
K.~L. Miller, J.~R. Males, O.~Guyon, L.~M. Close, D.~S. Doelman, F.~Snik, E.~H.
  Por, M.~J. Wilby, C.~U. Keller, C.~Bohlman \emph{et~al.}, \enquote{Spatial
  linear dark field control and holographic modal wavefront sensing with a vapp
  coronagraph on magao-x,} {\protect\JournalTitle{Journal of Astronomical
  Telescopes, Instruments, and Systems}} \textbf{5}, 049004 (2019).

\bibitem{Gonsalves1982}
R.~A. Gonsalves, \enquote{Phase retrieval and diversity in adaptive optics,}
  {\protect\JournalTitle{Optical Engineering}} \textbf{21}, 215829 (1982).

\bibitem{doelman2017}
D.~S. Doelman, F.~Snik, N.~Z. Warriner, and M.~J. Escuti, \enquote{{Patterned
  liquid-crystal optics for broadband coronagraphy and wavefront sensing},} in
  \emph{Techniques and Instrumentation for Detection of Exoplanets VIII,}  vol.
  10400 S.~Shaklan, ed., International Society for Optics and Photonics (SPIE,
  2017), pp. 224 -- 235.

\bibitem{miller2018development}
K.~L. Miller, \enquote{Development and demonstration of new focal plane
  wavefront sensing techniques for high-contrast direct imaging of exoplanets,}
  Ph.D. thesis, The University of Arizona (2018).

\bibitem{por2016}
E.~H. Por and C.~U. Keller, \enquote{Focal-plane electric field sensing with
  pupil-plane holograms,} in \emph{Adaptive Optics Systems V,}  vol. 9909
  (International Society for Optics and Photonics, 2016), p. 990959.

\bibitem{sun2019efficient}
H.~Sun, \enquote{Efficient wavefront sensing and control for space-based
  high-contrast imaging,} Ph.D. thesis, Princeton University (2019).

\bibitem{martinache2013asymmetric}
F.~Martinache, \enquote{The asymmetric pupil fourier wavefront sensor,}
  {\protect\JournalTitle{Publications of the Astronomical Society of the
  Pacific}} \textbf{125}, 422 (2013).

\bibitem{miller2017}
K.~Miller, O.~Guyon, and J.~Males, \enquote{Spatial linear dark field control:
  stabilizing deep contrast for exoplanet imaging using bright speckles,}
  {\protect\JournalTitle{Journal of Astronomical Telescopes, Instruments, and
  Systems}} \textbf{3}, 049002 (2017).

\bibitem{miller2020spatial}
K.~Miller, S.~Bos, J.~Lozi, O.~Guyon, D.~Doelman, S.~Vievard, A.~Sahoo, V.~Deo,
  N.~Jovanovic, F.~Martinache \emph{et~al.}, \enquote{Spatial linear dark field
  control on subaru/scexao,} {\protect\JournalTitle{Astronomy \& Astrophysics}}
   (2020).

\bibitem{bos2021submittedfirst}
S.~Bos, K.~Miller, J.~Lozi, O.~Guyon, D.~Doelman, S.~Vievard, A.~Sahoo, V.~Deo,
  N.~Jovanovic, F.~Martinache \emph{et~al.}, \enquote{First on-sky
  demonstration of spatial linear dark field control with the vector-apodizing
  phase plate at subaru/scexao,} {\protect\JournalTitle{Astronomy \&
  Astrophysics}}  (Submitted for publication).

\bibitem{escuti2016}
M.~J. Escuti, J.~Kim, and M.~W. Kudenov, \enquote{Controlling light with
  geometric-phase holograms,} {\protect\JournalTitle{Optics and Photonics
  News}} \textbf{27}, 22--29 (2016).

\bibitem{hornburg2014multiband}
K.~J. Hornburg, R.~K. Komanduri, and M.~J. Escuti, \enquote{Multiband
  retardation control using multi-twist retarders,} in \emph{Polarization:
  Measurement, Analysis, and Remote Sensing XI,}  vol. 9099 (International
  Society for Optics and Photonics, 2014), p. 90990Z.

\bibitem{hornburg2019highly}
K.~J. Hornburg, R.~K. Komanduri, and M.~J. Escuti, \enquote{Highly chromatic
  retardation via multi-twist liquid crystal films,}
  {\protect\JournalTitle{JOSA B}} \textbf{36}, D28--D33 (2019).

\bibitem{marois2006angular}
C.~Marois, D.~Lafreniere, R.~Doyon, B.~Macintosh, and D.~Nadeau,
  \enquote{Angular differential imaging: a powerful high-contrast imaging
  technique,} {\protect\JournalTitle{The Astrophysical Journal}} \textbf{641},
  556 (2006).

\bibitem{price2018astropy}
A.~M. Price-Whelan, B.~Sip{\H{o}}cz, H.~G{\"u}nther, P.~Lim, S.~Crawford,
  S.~Conseil, D.~Shupe, M.~Craig, N.~Dencheva, A.~Ginsburg \emph{et~al.},
  \enquote{The astropy project: Building an open-science project and status of
  the v2. 0 core package,} {\protect\JournalTitle{The Astronomical Journal}}
  \textbf{156}, 123 (2018).

\bibitem{gonzalez2017vip}
C.~A.~G. Gonzalez, O.~Wertz, O.~Absil, V.~Christiaens, D.~Defr{\`e}re,
  D.~Mawet, J.~Milli, P.-A. Absil, M.~Van~Droogenbroeck, F.~Cantalloube
  \emph{et~al.}, \enquote{Vip: Vortex image processing package for
  high-contrast direct imaging,} {\protect\JournalTitle{The Astronomical
  Journal}} \textbf{154}, 7 (2017).

\bibitem{amara2012pynpoint}
A.~Amara and S.~P. Quanz, \enquote{Pynpoint: an image processing package for
  finding exoplanets,} {\protect\JournalTitle{Monthly Notices of the Royal
  Astronomical Society}} \textbf{427}, 948--955 (2012).

\bibitem{stolker2019pynpoint}
T.~Stolker, M.~J. Bonse, S.~P. Quanz, A.~Amara, G.~Cugno, A.~J. Bohn, and
  A.~Boehle, \enquote{Pynpoint: a modular pipeline architecture for processing
  and analysis of high-contrast imaging data,} {\protect\JournalTitle{Astronomy
  \& Astrophysics}} \textbf{621}, A59 (2019).

\bibitem{wold1987principal}
S.~Wold, K.~Esbensen, and P.~Geladi, \enquote{Principal component analysis,}
  {\protect\JournalTitle{Chemometrics and intelligent laboratory systems}}
  \textbf{2}, 37--52 (1987).

\bibitem{samland2019high}
M.~S. Samland, \enquote{High-contrast imaging characterization of exoplanets,}
  Ph.D. thesis (2019).

\bibitem{close2010}
L.~M. Close, V.~Gasho, D.~Kopon, J.~Males, K.~B. Follette, K.~Brutlag,
  A.~Uomoto, and T.~Hare, \enquote{The magellan telescope adaptive secondary ao
  system: a visible and mid-ir ao facility,} in \emph{Adaptive Optics Systems
  II,}  vol. 7736 (International Society for Optics and Photonics, 2010), p.
  773605.

\bibitem{morzinski2014}
K.~M. Morzinski, L.~M. Close, J.~R. Males, D.~Kopon, P.~M. Hinz, S.~Esposito,
  A.~Riccardi, A.~Puglisi, E.~Pinna, R.~Briguglio \emph{et~al.},
  \enquote{Magao: Status and on-sky performance of the magellan adaptive optics
  system,} in \emph{Adaptive Optics Systems IV,}  vol. 9148 (International
  Society for Optics and Photonics, 2014), p. 914804.

\bibitem{jensen2017}
R.~Jensen-Clem, D.~Mawet, C.~A.~G. Gonzalez, O.~Absil, R.~Belikov, T.~Currie,
  M.~A. Kenworthy, C.~Marois, J.~Mazoyer, G.~Ruane \emph{et~al.}, \enquote{A
  new standard for assessing the performance of high contrast imaging systems,}
  {\protect\JournalTitle{The Astronomical Journal}} \textbf{155}, 19 (2017).

\bibitem{Sutlieff2019}
B.~J. Sutlieff, J.~L. Birkby, M.~A. Kenworthy, K.~M. Morzinski, J.~R. Males,
  A.~J. Bohn, D.~S. Doelman, and D.~Charbonneau, \enquote{{A vector Apodising
  Phase Plate view of an exoplanet atmosphere},} in \emph{Poster presented at
  Spirit of Lyot, Tokyo, Japan,}  (2019).

\bibitem{sutlieff2021}
B.~J. Sutlieff, A.~J. Bohn, J.~L. Birkby, M.~A. Kenworthy, K.~M. Morzinski,
  D.~S. Doelman, J.~R. Males, F.~Snik, L.~M. Close, P.~M. Hinz \emph{et~al.},
  \enquote{High-contrast observations of brown dwarf companion hr 2562 b with
  the vector apodizing phase plate coronagraph,} {\protect\JournalTitle{MNRAS}}
   (Submitted for publication).

\bibitem{haffert2016}
S.~Haffert, M.~Wilby, C.~Keller, and I.~Snellen, \enquote{The leiden exoplanet
  instrument (lexi): a high-contrast high-dispersion spectrograph,} in
  \emph{Ground-based and Airborne Instrumentation for Astronomy VI,}  vol. 9908
  (International Society for Optics and Photonics, 2016), p. 990867.

\bibitem{wilby2016}
M.~Wilby, C.~U. Keller, S.~Haffert, V.~Korkiakoski, F.~Snik, and A.~Pietrow,
  \enquote{Designing and testing the coronagraphic modal wavefront sensor: a
  fast non-common path error sensor for high-contrast imaging,} in
  \emph{Adaptive Optics Systems V,}  vol. 9909 (International Society for
  Optics and Photonics, 2016), p. 990921.

\bibitem{Jovanovic2015}
N.~Jovanovic, F.~Martinache, O.~Guyon, C.~Clergeon, G.~Singh, T.~Kudo,
  V.~Garrel, K.~Newman, D.~Doughty, J.~Lozi \emph{et~al.}, \enquote{The subaru
  coronagraphic extreme adaptive optics system: enabling high-contrast imaging
  on solar-system scales,} {\protect\JournalTitle{Publications of the
  Astronomical Society of the Pacific}} \textbf{127}, 890 (2015).

\bibitem{lozi2020status}
J.~Lozi, O.~Guyon, S.~Vievard, A.~Sahoo, V.~Deo, N.~Jovanovic, B.~Norris, M.-A.
  Martinod, B.~Mazin, A.~Walter \emph{et~al.}, \enquote{Status of the scexao
  instrument: recent technology upgrades and path to a system-level
  demonstrator for psi,} in \emph{Adaptive Optics Systems VII,}  vol. 11448
  (International Society for Optics and Photonics, 2020), p. 114480N.

\bibitem{groff2017first}
T.~Groff, J.~Chilcote, T.~Brandt, N.~J. Kasdin, M.~Galvin, C.~Loomis, M.~Rizzo,
  G.~Knapp, O.~Guyon, N.~Jovanovic \emph{et~al.}, \enquote{First light of the
  charis high-contrast integral-field spectrograph,} in \emph{Techniques and
  Instrumentation for Detection of Exoplanets VIII,}  vol. 10400 (International
  Society for Optics and Photonics, 2017), p. 1040016.

\bibitem{borgniet2019}
S.~Borgniet, A.-M. Lagrange, N.~Meunier, F.~Galland, L.~Arnold,
  N.~Astudillo-Defru, J.-L. Beuzit, I.~Boisse, X.~Bonfils, F.~Bouchy
  \emph{et~al.}, \enquote{Extrasolar planets and brown dwarfs around af-type
  stars-x. the sophie sample: combining the sophie and harps surveys to compute
  the close giant planet mass-period distribution around af-type stars,}
  {\protect\JournalTitle{Astronomy \& Astrophysics}} \textbf{621}, A87 (2019).

\bibitem{racine1999speckle}
R.~Racine, G.~A. Walker, D.~Nadeau, R.~Doyon, and C.~Marois, \enquote{Speckle
  noise and the detection of faint companions,}
  {\protect\JournalTitle{Publications of the Astronomical Society of the
  Pacific}} \textbf{111}, 587 (1999).

\bibitem{galicher2019}
R.~Galicher, P.~Baudoz, J.-R. Delorme, D.~Mawet, M.~Bottom, J.~K. Wallace,
  E.~Serabyn, and C.~Shelton, \enquote{Minimization of non-common path
  aberrations at the palomar telescope using a self-coherent camera,}
  {\protect\JournalTitle{Astronomy \& Astrophysics}} \textbf{631}, A143 (2019).

\bibitem{cote2018precursor}
O.~C{\^o}t{\'e}, G.~Allain, D.~Brousseau, M.-P. Lord, S.~Ouahbi, M.~Ouellet,
  D.~Patel, S.~Thibault, C.~Vall{\'e}e, R.~Belikov \emph{et~al.}, \enquote{A
  precursor mission to high contrast imaging balloon system,} in
  \emph{Ground-based and Airborne Instrumentation for Astronomy VII,}  vol.
  10702 (International Society for Optics and Photonics, 2018), p. 1070248.

\bibitem{thibault2019}
S.~Thibault, G.~Allain, O.~C{\^o}t{\'e}, M.~Ouellet, D.~Patel, and
  C.~Vall{\'e}e, \enquote{Stringent and result-oriented training requirements
  at the heart of research funding opportunities: the case of the csa fast
  funding activity and the hicibas project,} in \emph{Education and Training in
  Optics and Photonics,}  (Optical Society of America, 2019), p. 11143\_130.

\bibitem{allain2018}
G.~Allain, D.~Brousseau, S.~Thibault, C.~Vall{\'e}e, M.~Ouellet, J.-P.
  V{\'e}ran, and O.~Daigle, \enquote{First on-sky results, performance, and
  future of the hicibas-lowfs,} in \emph{Adaptive Optics Systems VI,}  vol.
  10703 (International Society for Optics and Photonics, 2018), p. 107035T.

\bibitem{males2018}
J.~R. Males, L.~M. Close, K.~Miller, L.~Schatz, D.~Doelman, J.~Lumbres,
  F.~Snik, A.~Rodack, J.~Knight, K.~Van~Gorkom \emph{et~al.}, \enquote{Magao-x:
  project status and first laboratory results,} in \emph{Adaptive Optics
  Systems VI,}  vol. 10703 (International Society for Optics and Photonics,
  2018), p. 1070309.

\bibitem{males2020magao}
J.~R. Males, L.~M. Close, O.~Guyon, A.~D. Hedglen, K.~Van~Gorkom, J.~D. Long,
  M.~Kautz, J.~Lumbres, L.~Schatz, A.~Rodack \emph{et~al.}, \enquote{Magao-x
  first light,} in \emph{Adaptive Optics Systems VII,}  vol. 11448
  (International Society for Optics and Photonics, 2020), p. 114484L.

\bibitem{close2020separation}
L.~M. Close, \enquote{The separation and h$\alpha$ contrasts of massive
  accreting planets in the gaps of transitional disks: Predicted h$\alpha$
  protoplanet yields for adaptive optics surveys,} {\protect\JournalTitle{The
  Astronomical Journal}} \textbf{160}, 221 (2020).

\bibitem{close2020prediction}
L.~M. Close, J.~Males, J.~D. Long, K.~Van~Gorkom, A.~D. Hedglen, M.~Kautz,
  J.~Lumbres, S.~Haffert, K.~Follette, K.~Wagner \emph{et~al.},
  \enquote{Prediction of the planet yield of the maxprotoplanets high-contrast
  survey for h-alpha protoplanets with magao-x based on first light contrasts,}
  in \emph{Adaptive Optics Systems VII,}  vol. 11448 (International Society for
  Optics and Photonics, 2020), p. 114480U.

\bibitem{skrutskie2010}
M.~Skrutskie, T.~Jones, P.~Hinz, P.~Garnavich, J.~Wilson, M.~Nelson,
  E.~Solheid, O.~Durney, W.~Hoffmann, V.~Vaitheeswaran \emph{et~al.},
  \enquote{The large binocular telescope mid-infrared camera (lmircam): final
  design and status,} in \emph{Ground-based and Airborne Instrumentation for
  Astronomy III,}  vol. 7735 (International Society for Optics and Photonics,
  2010), p. 77353H.

\bibitem{skemer2015first}
A.~J. Skemer, P.~Hinz, M.~Montoya, M.~F. Skrutskie, J.~Leisenring, O.~Durney,
  C.~E. Woodward, J.~Wilson, M.~Nelson, V.~Bailey \emph{et~al.}, \enquote{First
  light with ales: A 2-5 micron adaptive optics integral field spectrograph for
  the lbt,} in \emph{Techniques and Instrumentation for Detection of Exoplanets
  VII,}  vol. 9605 (International Society for Optics and Photonics, 2015), p.
  96051D.

\bibitem{skemer2018ales}
A.~J. Skemer, P.~Hinz, J.~Stone, M.~Skrutskie, C.~E. Woodward, J.~Leisenring,
  and Z.~Briesemeister, \enquote{{ALES: overview and upgrades},} in
  \emph{Ground-based and Airborne Instrumentation for Astronomy VII,}  vol.
  10702 C.~J. Evans, L.~Simard, and H.~Takami, eds., International Society for
  Optics and Photonics (SPIE, 2018), pp. 78 -- 85.

\bibitem{amico2012design}
P.~Amico, E.~Marchetti, F.~Pedichini, A.~Baruffolo, B.~Delabre, M.~Duchateau,
  M.~Ekinci, D.~Fantinel, E.~Fedrigo, G.~Finger \emph{et~al.}, \enquote{The
  design of eris for the vlt,} in \emph{Ground-based and Airborne
  Instrumentation for Astronomy IV,}  vol. 8446 (International Society for
  Optics and Photonics, 2012), p. 844620.

\bibitem{davies2018eris}
R.~Davies, S.~Esposito, H.-M. Schmid, W.~Taylor, G.~Agapito, A.~A. Berbel,
  A.~Baruffolo, V.~Biliotti, B.~Biller, M.~Black \emph{et~al.}, \enquote{Eris:
  revitalising an adaptive optics instrument for the vlt,} in
  \emph{Ground-based and Airborne Instrumentation for Astronomy VII,}  vol.
  10702 (International Society for Optics and Photonics, 2018), p. 1070209.

\bibitem{kenworthy2018high}
M.~A. Kenworthy, F.~Snik, C.~U. Keller, D.~Doelman, E.~H. Por, O.~Absil,
  B.~Carlomagno, M.~Karlsson, E.~Huby, A.~M. Glauser \emph{et~al.},
  \enquote{High contrast imaging for the enhanced resolution imager and
  spectrometer (eris),} in \emph{Ground-based and Airborne Instrumentation for
  Astronomy VII,}  vol. 10702 (International Society for Optics and Photonics,
  2018), p. 1070246.

\bibitem{boehle2018cryogenic}
A.~Boehle, A.~M. Glauser, M.~A. Kenworthy, F.~Snik, D.~Doelman, S.~P. Quanz,
  and M.~R. Meyer, \enquote{Cryogenic characterization of the grating vector
  app coronagraph for the upcoming eris instrument at the vlt,} in
  \emph{Ground-based and Airborne Instrumentation for Astronomy VII,}  vol.
  10702 (International Society for Optics and Photonics, 2018), p. 107023Y.

\bibitem{boehle2021submitted}
A.~Boehle, D.~Doelman \emph{et~al.}, \enquote{Cryogenic characterization of the
  grating vector app for eris,} {\protect\JournalTitle{Journal of Astronomical
  Telescopes, Instruments, and Systems}}  (Submitted for publication).

\bibitem{gilmozzi2007european}
R.~Gilmozzi and J.~Spyromilio, \enquote{The european extremely large telescope
  (e-elt),} {\protect\JournalTitle{The Messenger}} \textbf{127}, 3 (2007).

\bibitem{brandl2016status}
B.~R. Brandl, T.~Ag{\'o}cs, G.~Aitink-Kroes, T.~Bertram, F.~Bettonvil, R.~van
  Boekel, O.~Boulade, M.~Feldt, A.~Glasse, A.~Glauser \emph{et~al.},
  \enquote{Status of the mid-infrared e-elt imager and spectrograph metis,} in
  \emph{Ground-based and Airborne Instrumentation for Astronomy VI,}  vol. 9908
  (International Society for Optics and Photonics, 2016), p. 990820.

\bibitem{davies2016micado}
R.~Davies, J.~Schubert, M.~Hartl, J.~Alves, Y.~Cl{\'e}net, F.~Lang-Bardl,
  H.~Nicklas, J.-U. Pott, R.~Ragazzoni, E.~Tolstoy \emph{et~al.},
  \enquote{Micado: first light imager for the e-elt,} in \emph{Ground-based and
  Airborne Instrumentation for Astronomy VI,}  vol. 9908 (International Society
  for Optics and Photonics, 2016), p. 99081Z.

\bibitem{quanz2015direct}
S.~P. Quanz, I.~Crossfield, M.~R. Meyer, E.~Schmalzl, and J.~Held,
  \enquote{Direct detection of exoplanets in the 3--10 $\mu$m range with
  e-elt/metis,} {\protect\JournalTitle{International Journal of Astrobiology}}
  \textbf{14}, 279--289 (2015).

\bibitem{kenworthy2018review}
M.~A. Kenworthy, O.~Absil, B.~Carlomagno, T.~Ag{\'o}cs, E.~H. Por, S.~Bos,
  B.~Brandl, and F.~Snik, \enquote{A review of high contrast imaging modes for
  metis,} in \emph{Ground-based and Airborne Instrumentation for Astronomy
  VII,}  vol. 10702 (International Society for Optics and Photonics, 2018), p.
  10702A3.

\bibitem{gluck2016simulation}
M.~Gl{\"u}ck, J.-U. Pott, and O.~Sawodny, \enquote{Simulation of an
  accelerometer-based feedforward vibration suppression in an adaptive optics
  system for micado,} in \emph{Adaptive Optics Systems V,}  vol. 9909
  (International Society for Optics and Photonics, 2016), p. 99093N.

\bibitem{davies2018}
Y.~{Cl{\'e}net}, T.~{Buey}, E.~{Gendron}, Z.~{Hubert}, F.~{Vidal}, M.~{Cohen},
  F.~{Chapron}, A.~{Sevin}, P.~{F{\'e}dou}, G.~{Barbary}, P.~{Baudoz},
  B.~{Borgo}, S.~{Ben Nejma}, V.~{Chambouleyron}, V.~{D{\'e}o}, O.~{Dupuis},
  S.~{Durand}, F.~{Ferreira}, J.~{Gaudemard}, D.~{Gratadour}, E.~{Huby}, J.-M.
  {Huet}, B.~{Le Ruyet}, N.~{Nguyen-Tuong}, C.~{Perrot}, S.~{Thijs},
  Y.~{Youn{\`e}s}, G.~{Rousset}, P.~{Feautrier}, G.~{Zins}, E.~{Diolaiti},
  P.~{Ciliegi}, S.~{Esposito}, L.~{Busoni}, J.~{Schubert}, M.~{Hartl},
  V.~{H{\"o}rmann}, and R.~{Davies}, \enquote{{The MICADO first-light imager
  for the ELT: towards the preliminary design review of the MICADO-MAORY
  SCAO},} in \emph{Adaptive Optics Systems VI,}  vol. 10703 of \emph{Society of
  Photo-Optical Instrumentation Engineers (SPIE) Conference Series} L.~M.
  {Close}, L.~{Schreiber}, and D.~{Schmidt}, eds. (2018), p. 1070313.

\bibitem{clenet2018}
Y.~{Cl{\'e}net}, T.~{Buey}, E.~{Gendron}, Z.~{Hubert}, F.~{Vidal}, M.~{Cohen},
  F.~{Chapron}, A.~{Sevin}, P.~{F{\'e}dou}, G.~{Barbary}, P.~{Baudoz},
  B.~{Borgo}, S.~{Ben Nejma}, V.~{Chambouleyron}, V.~{D{\'e}o}, O.~{Dupuis},
  S.~{Durand}, F.~{Ferreira}, J.~{Gaudemard}, D.~{Gratadour}, E.~{Huby}, J.-M.
  {Huet}, B.~{Le Ruyet}, N.~{Nguyen-Tuong}, C.~{Perrot}, S.~{Thijs},
  Y.~{Youn{\`e}s}, G.~{Rousset}, P.~{Feautrier}, G.~{Zins}, E.~{Diolaiti},
  P.~{Ciliegi}, S.~{Esposito}, L.~{Busoni}, J.~{Schubert}, M.~{Hartl},
  V.~{H{\"o}rmann}, and R.~{Davies}, \enquote{{The MICADO first-light imager
  for the ELT: towards the preliminary design review of the MICADO-MAORY
  SCAO},} in \emph{Adaptive Optics Systems VI,}  vol. 10703 of \emph{Society of
  Photo-Optical Instrumentation Engineers (SPIE) Conference Series} L.~M.
  {Close}, L.~{Schreiber}, and D.~{Schmidt}, eds. (2018), p. 1070313.

\bibitem{diolaiti2016maory}
E.~Diolaiti, P.~Ciliegi, R.~Abicca, G.~Agapito, C.~Arcidiacono, A.~Baruffolo,
  M.~Bellazzini, V.~Biliotti, M.~Bonaglia, G.~Bregoli \emph{et~al.},
  \enquote{Maory: adaptive optics module for the e-elt,} in \emph{Adaptive
  Optics Systems V,}  vol. 9909 (International Society for Optics and
  Photonics, 2016), p. 99092D.

\bibitem{baudoz2019}
P.~{Baudoz}, E.~{Huby}, and Y.~{Cl{\'e}net}, \enquote{{Exoplanetary systems
  study with MICADO},} in \emph{SF2A-2019: Proceedings of the Annual meeting of
  the French Society of Astronomy and Astrophysics,}  P.~{Di Matteo},
  O.~{Creevey}, A.~{Crida}, G.~{Kordopatis}, J.~{Malzac}, J.~B. {Marquette},
  M.~{N'Diaye}, and O.~{Venot}, eds. (2019), p.~Di.

\bibitem{perrot2018}
C.~{Perrot}, P.~{Baudoz}, A.~{Boccaletti}, G.~{Rousset}, E.~{Huby},
  Y.~{Cl{\'e}net}, S.~{Durand}, and R.~{Davies}, \enquote{{Design study and
  first performance simulation of the ELT/MICADO focal plane coronagraphs},}
  {\protect\JournalTitle{arXiv e-prints}} arXiv:1804.01371 (2018).

\bibitem{Lacour2014}
S.~{Lacour}, P.~{Baudoz}, E.~{Gendron}, A.~{Boccaletti}, R.~{Galicher},
  Y.~{Cl{\'e}net}, D.~{Gratadour}, T.~{Buey}, G.~{Rousset}, M.~{Hartl}, and
  R.~{Davies}, \enquote{{An aperture masking mode for the MICADO instrument},}
  in \emph{Ground-based and Airborne Instrumentation for Astronomy V,}  vol.
  9147 of \emph{Society of Photo-Optical Instrumentation Engineers (SPIE)
  Conference Series} S.~K. {Ramsay}, I.~S. {McLean}, and H.~{Takami}, eds.
  (2014), p. 91479F.

\bibitem{snik2014combining}
F.~Snik, G.~Otten, M.~Kenworthy, D.~Mawet, and M.~Escuti, \enquote{Combining
  vector-phase coronagraphy with dual-beam polarimetry,} in \emph{Ground-based
  and Airborne Instrumentation for Astronomy V,}  vol. 9147 (International
  Society for Optics and Photonics, 2014), p. 91477U.

\bibitem{kuhn2001imaging}
J.~Kuhn, D.~Potter, and B.~Parise, \enquote{Imaging polarimetric observations
  of a new circumstellar disk system,} {\protect\JournalTitle{The Astrophysical
  Journal Letters}} \textbf{553}, L189 (2001).

\bibitem{doelman2019multi}
D.~S. Doelman, M.~J. Escuti, and F.~Snik, \enquote{Multi-color holography with
  a two-stage patterned liquid-crystal element,} {\protect\JournalTitle{Optical
  Materials Express}} \textbf{9}, 1246--1256 (2019).

\end{thebibliography}

\end{document}